\documentstyle[11pt,epic]{article}
\newcommand{\ol}{\overline}
\newcommand{\ul}{\underline}
\newcommand{\mod}{\hbox{\mod }}

\begin{document}

\title{Some Linear-Time Algorithms for Systolic Arrays%
\thanks{Invited paper presented by the first author at the Ninth World
Computer Congress (Paris, September 1983)
and published in {\em Information Processing~83},
R.~E.~A.~Mason (ed.), Elsevier Science Publishers B.~V.~(North Holland),
865--876.
Retyped (with corrections) in \LaTeX\ by Frances Page, October 2000.}
}

\author{Richard P.~Brent
\thanks{R.~P.~Brent is with the 
Mathematical Sciences Institute,
Australian National University, Canberra, Australia.}
\and
H.~T.~Kung
\thanks{H.~T.~Kung is with the Department of Computer Science,
Carnegie-Mellon University, Pittsburgh, PA 15213, USA.
(Current address: Department of Computer Science, Harvard University)}
\and
Franklin T.~Luk
\thanks{F.~T.~Luk is with the Department of Computer Science,
Cornell University, Ithaca, New York 14853, USA.
(Current address: Hong Kong Baptist University)\hspace*{10em}
\hbox{Copyright \copyright\ 1983--2010, the authors.}
\hspace*{\fill}
rpb079 typeset using \LaTeX.
}
}

\date{}

\maketitle

\begin{abstract} %
\noindent
We survey some recent results on linear-time algorithms for systolic
arrays.  In particular, we show how the greatest common divisor (GCD)
of two polynomials of degree $n$ over a finite field can be computed
in time $O(n)$ on a linear systolic array of $O(n)$ cells; similarly
for the GCD of two $n$-bit binary numbers.  We show how $n$ by $n$
Toeplitz systems of linear equations can be solved in time $O(n)$
on a linear array of $O(n)$ cells, each of which has constant memory
size (independent of $n$).  Finally, we outline how a two-dimensional
square array of $O(n)$ by $O(n)$ cells can be used to solve (to
working accuracy) the eigenvalue problem for a symmetric real $n$
by $n$ matrix in time $O(nS(n))$.  Here $S(n)$ is a slowly growing
function of $n$; for practical purposes $S(n)$ can be regarded as a
constant.  In addition to their theoretical interest, these results
have potential applications in the areas of error-correcting codes,
symbolic and algebraic computation, signal processing and image
processing.  For example, systolic GCD arrays for error correction
have been implemented with the microprogrammable ``PSC'' chip.
\end{abstract}

\section{Introduction}
\thispagestyle{empty}

A ``systolic array'' is a regular array of simple machines or ``cells''
with a nearest-neighbour interconnection pattern. A pipeline is an
example of a linear systolic array in which data flows only in one direction,
but systolic arrays may be two-dimensional (rectangular, triangular or
hexagonal) and data may flow between the cells in several different directions
and at several different speeds.  The concept of systolic arrays has recently 
been developed by H.T.~Kung and his students [24, 35, 36, 37, 38, 43],
although related ideas can be found in earlier work on models of computation
[19, 29].\\

Systolic arrays may be implemented as synchronous or asynchronous systems,
but for expository purposes we shall consider only synchronous systems.
Systolic arrays are not necessarily fixed, special-purpose systems; they
can be programmed [5, 21, 49, 57] or simulated by more general parallel
machines [34, 55], although at some loss of efficiency.\\

A ``systolic algorithm'' is a specification of the operation of each cell
in a systolic array, together with a specification of the interconnection
pattern of the array.  Systolic algorithms have been suggested for solving
many compute-bound problems, e.g.~binary and polynomial arithmetic,
convolution, filtering, matrix multiplication, solution of linear systems
and least squares problems, and geometric problems [6, 18, 25, 37, 39].
Here we survey some recent results on systolic algorithms.  The results
are interesting because they show that systolic arrays can be used to solve
certain important problems in linear (or almost linear) time; the problems
considered have practical applications in areas such as error correction, 
symbolic computation, signal processing and image processing.\\

The problems considered here are the computation of greatest common divisiors 
of polynomials (over a finite field) and of binary integers, the solution of
Toeplitz systems of linear equations, and the solution of the symmetric 
eigenvalue problem.  The first two problems require a linear array with 
uni-directional data flow (i.e.~a pipeline), the third requires a linear
array with bi-directional data flow, and the fourth requires a square
(two-dimensional) array.  The third and fourth problems require the use of 
floating-point arithmetic, and the fourth requires an iterative rather than
a direct solution.  The third and fourth problems also illustrate a common 
technique for converting a ``semi-systolic'' array (i.e.~one with global 
broadcasting) into a true systolic array [43].  Because of space limitations we 
have had to omit many details, for which we refer the reader to the original
papers [10, 11, 14, 15].

\pagebreak[3]
\section{Polynomial GCD computation}

The polynomial GCD problem is to compute a greatest common divisor of any two 
nonzero polynomials.  This problem is fundamental to algebraic and symbolic
computations and to the decoder implemetations for a variety of 
error-correcting codes [9, 32, 46].  Many algorithms for solving the GCD
problem are known [2, 7, 32].  However, for direct hardware implementation these
algorithms are too irregular and/or too complex to be useful.  For example,
the classical Euclidean algorithm involves a sequence of divisions of
polynomials whose size can only be determined during the computation.  We
shall describe some simple and regular systolic structures which can provide 
efficient hardware solutions to the GCD problem.\\
 
In particular, we describe
a systolic array of $m+n+1$ cells which can find a GCD of any two
polynomials of degrees $m$ and $n\,$.  Figure~1 illustrates that the
coefficients of the given polynomials
${\displaystyle\sum^n_{i=0}}~a_i x^i$ and ${\displaystyle\sum^m_{j=0}}~b_j x^j$
enter the leftmost cell and the output (their GCD) emerges from the rightmost
cell of the array.  \\

More precisely, if a unit of time is taken to be the cell
cycle time (which is essentially the time required to perform a division or a
multiplication and an addition), the $2(m+n+1)$ time units after $a_n$ and
$b_m$ enter the leftmost cell, the co-efficients of the GCD start emerging
from the rightmost cell at the rate of one co-efficient per unit time.
Unlike the systolic arrays described in Sections 4 and 5, the array
illustrated in Figure~1 is a pipeline, as data flows through it in only
one direction (although not necessarily at constant speed).

\begin{center}
\setlength{\unitlength}{1mm}
\begin{picture}(105,15)(0,0)

\small

\put(0,7.5){\makebox(5,5){$a_0$}}
\put(0,7.5){\vector(1,0){5}}
\put(10,7.5){\makebox(5,5){$a_1$}}
\put(10,7.5){\vector(1,0){5}}
\put(10,2.5){\makebox(5,5){$b_0$}}
\put(10,2.5){\vector(1,0){5}}
\multiput(19,7.5)(3,0){3}{\bf .}
\multiput(19,2.5)(3,0){3}{\bf .}
\put(30,7.5){\makebox(5,5){$a_n$}}
\put(30,7.5){\vector(1,0){5}}
\put(30,2.5){\makebox(5,5){$b_m$}}
\put(30,2.5){\vector(1,0){5}}
\put(35,0){\framebox(5,10){}}
\put(40,7.5){\vector(1,0){5}}
\put(40,2.5){\vector(1,0){5}}
\put(45,0){\framebox(5,10){}}
\put(50,7.5){\vector(1,0){5}}
\put(50,2.5){\vector(1,0){5}}
\multiput(64,7.5)(3,0){3}{\bf .}
\multiput(64,2.5)(3,0){3}{\bf .}
\put(80,7.5){\vector(1,0){5}}
\put(80,2.5){\vector(1,0){5}}
\put(85,0){\framebox(5,10){}}
\put(90,7.5){\vector(1,0){5}}
\put(90,2.5){\vector(1,0){5}}
\put(95,5){\makebox(10,5){\rm GCD}}
\put(95,0){\makebox(5,5){0}}
\put(50,12.5){\vector(-1,0){15}}
\put(50,10){\makebox(25,5){$m+n+1$ cells}}
\put(75,12.5){\vector(1,0){15}}

\end{picture}
\vspace*{3mm}

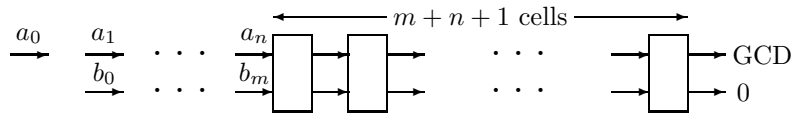
\ul{Figure~1: Systolic array for polynomial GCD}
\end{center}

The systolic arrays described in this paper are suitable for VLSI implementation 
[47] and can achieve high throughputs.  The systolic polynomial GCD algorithms
were developed in order to implement a decoder for Reed-Solomon 
error-correcting codes with the Programmable Systolic Chip (PSC) [21].\\

Since it is not easy to understand some of the more complicated systolic algorithms, we 
shall start with the basic ideas and describe some simple algorithms first.
Hopefully informal arguments will convince the reader that our algorithms
are correct.  Formal correctness proofs are beyond the scope of this paper.
Nevertheless, every systolic algorithm mentioned below has been tested by
simulation, using Pascal or Lisp programs on a serial computer, so we may
have some degree of confidence in their correctness.

\subsection{GCD-preserving transformations}

All well-known algorithms for solving the polynomial GCD problem are based
on the general technique of reducing the degrees of the two given polynomials
by ``GCD-preserving'' transformations.  A GCD-preserving transformation
transforms a pair $(A, B)$ of polynomials into another pair $(\ol{A}, \ol{B})$
such that a GCD of $A$ and $B$ is also a GCD of $\ol{A}$ and $\ol{B}$, and
vice versa.  (We say ``a GCD'' because a GCD over a finite field is not
generally unique.)  When one of the two polynomials is reduced to zero by
a sequence of such transformations, the other polynomial will be a GCD of the
original two polynomials.  We use this general technique, but choose very
simple GCD-preserving transformations to permit their implementation by a 
systolic array.\\

We assume throughout this section that the co-efficients of the polynomials
belong to a finite field.  This is true for the decoder application for
error-correcting codes; in [10] it is shown that straightforward modifications
of our designs require no divisions and work over any unique factorisation
domain.  We define two GCD-preserving transformations, $R_A$ and $R_B\,$.
Let\linebreak 
$A= a_ix^i + \cdots + a_0$ and $B = b_jx^j + \cdots b_0$ be the two
polynomials to be transformed, where $a_i \neq 0$ and $b_j \neq 0\,$.

\begin{center}
{\bf Transformation} {\boldmath$R_A$} (for the case $i-j\geq 0)\,$:\\[10pt]

\setlength{\unitlength}{1mm}
\begin{picture}(65,15)(0,0)

\small
\put(0,7.5){\makebox(5,5){$A$}}
\put(5,10){\vector(1,0){5}}
\put(0,2.5){\makebox(5,5){$B$}}
\put(5, 5){\vector(1,0){5}}
\put(10,0){\dashbox(20,15){$R_A$}}
\put(30,10){\vector(1,0){5}}
\put(37,7.5){\makebox(40,5)[l]{$\overline{A}=A-qx^dB$
				where $d=i-j$ and $a_i/b_j\,$.}}
\put(30,5){\vector(1,0){5}}
\put(37,2.5){\makebox(40,5)[l]{$\overline{B}=B$}}
\end{picture}\\
\end{center}

\begin{center}
{\bf Transformation} {\boldmath$R_B$} (for the case $i-j < 0)\,$:\\[10pt]

\setlength{\unitlength}{1mm}
\begin{picture}(65,15)(0,0)

\small
\put(0,7.5){\makebox(5,5){$A$}}
\put(5,10){\vector(1,0){5}}
\put(0,2.5){\makebox(5,5){$B$}}
\put(5, 5){\vector(1,0){5}}
\put(10,0){\dashbox(20,15){$R_B$}}
\put(30,10){\vector(1,0){5}}
\put(37,7.5){\makebox(40,5)[l]{$\overline{A}=A$}}
\put(30,5){\vector(1,0){5}}
\put(37,2.5){\makebox(40,5)[l]{$\overline{B}=B-qx^dA$
				where $d=j-i$ and $q=b_j/a_i\,$.}}
\end{picture}\\
\end{center}

It is obvious that both the transformations are GCD-preserving.  Furthermore,
$R_A$ decreases the degree of $A\,$, i.e.~$\deg \ol{A} < \deg A\,$, and
$R_B$ decreases the degree of $B\,$, i.e.~$\deg \ol{B} < \deg B\,$.
(For notational convenience we assume that the degree of the zero polynomial
is $-1\,$.)

\subsection{Transformation sequence for polynomial GCD computation}

To compute a GCD of two given polynomials $A_0$ and $B_0$ of degrees $n$ and
$m\,$, we can apply a sequence of GCD-preserving transformations, each one
being either $R_A$ or $R_B\,$, until one of the two polynomials is transformed
to zero;  at this point the other (nonzero) polynomial is a GCD of $A_0$
and $B_0\,$.  We call this sequence of transformations the {\bf transformation
sequence} for $A_0$\pagebreak[4] 
and $B_0\,$, and denote it by $(T_1, T_2, \dots , T_k)$
for some $k\,$.  $T_i$ transforms $(A_{i-1}, B_{i-1})$ to $(A_i, B_i)\,$.
Note that the transformation sequence is uniquely defined for given $A_0$
and $B_0\,$.\\

An instructive way to view the function of the transformation sequence is to
imagine that polynomials $A_0$ and $B_0$ move through the transformation
``stages'' $T_1, T_2, \dots, T_k$ from left to right, being transformed
at each stage; when they emerge from the last stage $T_k\,$, one will be the
zero polynomial and the other will be a GCD of $A_0$ and $B_0\,$.\\

Suppose that transformation $T_i$ reduces the sum of the degrees of its input
polynomials $A_{i-1}$ and $B_{i-1}$ by $\delta_i > 0\,$.  We call $\delta_i$
the {\bf reduction value} of $T_i\,$. Since the sum of the degrees of $A_0$
and $B_0$ at the beginning of the GCD is $n+m\,$, we have
${\displaystyle\sum^k_{i=1}}~\delta_i \leq n+m+1\,$.

\subsection{A systolic array for polynomial GCD computation}

We now specify a systolic array of $n+m+1$ cells which can compute a GCD
of any two input polynomials $A_0$ and $B_0$ (not both zero) of degrees no
more than $n$ and $m\,$, respectively.\\

Consider the transformation sequence $(T_1, \dots , T_k)$ for $A_0$ and
$B_0\,$.  For each $i = 1, \dots , k\,$, transformation $T_i$ can be
realised by a subarray of $\delta_i$ cells, where $\delta_i$ is the reduction 
value of $T_i\,$.\\

Since ${\displaystyle\sum^k_{i=1}}~\delta_i \leq n+m+1\,$, a systolic array
with $n+m+1$ calls can realise all the transformations.  This is illustrated 
in Figure~2.

\begin{center}

\setlength{\unitlength}{1mm}
\begin{picture}(85,15)(0,0)       
\small
\put(10,0){\dashbox(15,9){$T_1$}}
\put(35,0){\dashbox(15,9){$T_2$}}
\put(60,0){\dashbox(15,9){$T_3$}}
\put(0,6){\vector(1,0){10}} \put(0,6.5){\makebox(10,3){$A_0$}}
\put(0,3){\vector(1,0){10}} \put(0,0){\makebox(10,2){$B_0$}}
\put(25,6){\vector(1,0){10}} \put(25,6.5){\makebox(10,3){$A_1$}}
\put(25,3){\vector(1,0){10}} \put(25,0){\makebox(10,2){$B_1$}}
\put(50,6){\vector(1,0){10}} \put(50,6.5){\makebox(10,3){$A_2$}}
\put(50,3){\vector(1,0){10}} \put(50,0){\makebox(10,2){$B_2$}}
\put(75,6){\vector(1,0){10}} \put(75,6.5){\makebox(10,3){$A_3$}}
\put(75,3){\vector(1,0){10}} \put(75,0){\makebox(10,2){$B_3$}}
\put(10,9){\makebox(15,6){$\delta_1 = 2$}}
\put(35,9){\makebox(15,6){$\delta_2 = 3$}}
\put(60,9){\makebox(15,6){$\delta_3 = 2$}}
\end{picture}

\vspace*{6mm}

\setlength{\unitlength}{1mm}
\begin{picture}(95,25)(0,0)       
\small
\put(10,10){\framebox(5,9){}} \put(20,10){\framebox(5,9){}}
\put(7.5,7.5){\dashbox(20,14){}}
\put(10,1){\makebox(15,5){2 cells for $T_1$}}
\put(2,15.5){\makebox(5,5){$A_0$}} \put(2,8.5){\makebox(5,5){$B_0$}}

\put(35,10){\framebox(5,9){}} \put(45,10){\framebox(5,9){}}
\put(55,10){\framebox(5,9){}}
\put(32.5,7.5){\dashbox(30,14){}} 
\put(35,1){\makebox(25,5){3 cells for $T_2$}}
\put(27.5,15.5){\makebox(5,5){$A_1$}} \put(27.5,8.5){\makebox(5,5){$B_1$}}

\put(70,10){\framebox(5,9){}} \put(80,10){\framebox(5,9){}}
\put(67.5,7.5){\dashbox(20,14){}}
\put(70,1){\makebox(15,5){2 cells for $T_3$}}
\put(62.5,15.5){\makebox(5,5){$A_2$}} \put(62.5,8.5){\makebox(5,5){$B_2$}}

\put(0,16){\vector(1,0){10}} \put(0,13){\vector(1,0){10}}
\put(15,16){\vector(1,0){5}} \put(15,13){\vector(1,0){5}}
\put(25,16){\vector(1,0){10}} \put(25,13){\vector(1,0){10}}
\put(40,16){\vector(1,0){5}} \put(40,13){\vector(1,0){5}}
\put(50,16){\vector(1,0){5}} \put(50,13){\vector(1,0){5}}
\put(60,16){\vector(1,0){10}} \put(60,13){\vector(1,0){10}}
\put(75,16){\vector(1,0){5}} \put(75,13){\vector(1,0){5}}
\put(85,16){\vector(1,0){10}} \put(85,13){\vector(1,0){10}}
\put(88.5,15.5){\makebox(5,5){$A_3$}} \put(88.5,8.5){\makebox(5,5){$B_3$}}

\end{picture}

\vspace*{3mm}
{\begin{tabular}{ll}
\hspace*{-1em}
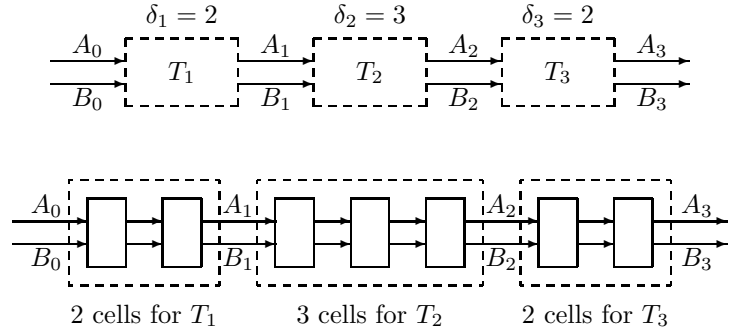
Figure~2: & (a) Transformation sequence, and\\
          & (b) its realisation by three concatenated 
                                        systolic subarrays
\hspace*{-0.8em}\\ \hline
\end{tabular}}

\end{center}

\subsubsection{The basic idea for realising a single transformation}

Let $T$ be any transformation in the transformation sequence $(T_1, \dots ,
T_k)\,$, and~$\delta$ its reduction value.  We illustrate how a subarray
with $\delta$ cells can realise $T\,$, assuming that we know which of
$R_A$ and $R_B$ the transformation $T$ is (see Section~2.3.2 below). We
consider the case when $T$ is $R_A\,$; the case when $T$ is $R_B$ can be
treated similarly.  Without loss of generality, we can assume that $T$
transforms $(A,B)$ to $(\ol{A}, \ol{B}) = (A-qx^d B, B)\,$ where
\medskip

\hspace*{46mm}
\begin{tabular}{l}
$A = a_ix^i + \cdots + a_0\,$, $B = b_jx^j + \cdots + b_0\,$, \\
$a_i \neq 0\,$, $b_j \neq 0\,$,
$q = a_i/b_j\,$, and
$d = i-j \geq 0\,$.\\
\end{tabular}\\

Note that either $\ol{A} = 0$ or 
$\ol{A} = \ol{a}_{i-\delta}x^{i-\delta} + \cdots + \ol{a}_0\,$, 
where $\ol{a}_{i-\delta} \neq 0\,$.  The systolic subarray for 
realising $T$ is shown in Figure~3.\\

Terms of $A$ and $B$ move through the subarray in a serial manner, high degree
terms first (there is a dual with low degree terms first).  The nonzero
leading terms of $A$ and $B$ are aligned so that they enter the leftmost
cell of the subarray during the same cycle.  Besides the systolic data paths
for $a$ and $b\,$, there is a 1-bit wide systolic control path,  denoted
by {\tt start}; a true (i.e.~1) value on this path signals to a cell the 
beginning of a new GCD computation in the following cycle.  In Figure~3
and below, 1-bit wide systolic control paths and associated registers are 
shown by dotted arrows and boxes.\\

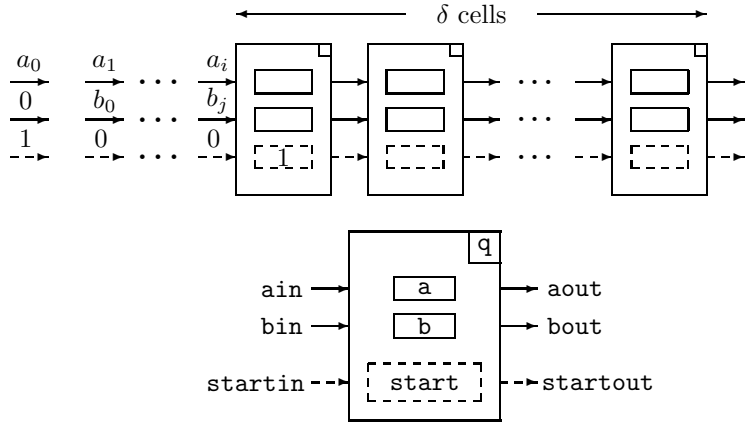
\begin{center}
\setlength{\unitlength}{1mm}
\begin{picture}(100,55)(0,0)       
\small
\put(0,45){\vector(1,0){5}} \put(0,45){\makebox(5,5){$a_0$}}
\put(0,40){\vector(1,0){5}} \put(-.5,40){\makebox(5,5){$0$}}
\multiput(0,35)(2,0){2}{\line(1,0){1}} \put(4,35){\vector(1,0){1}} 
\put(-.5,35){\makebox(5,5){$1$}}

\put(10,45){\vector(1,0){5}} \put(10,45){\makebox(5,5){$a_1$}}
\put(10,40){\vector(1,0){5}} \put(10,40){\makebox(5,5){$b_0$}}
\multiput(10,35)(2,0){2}{\line(1,0){1}}
\put(14,35){\vector(1,0){1}} \put(9.5,35){\makebox(5,5){$0$}}

\put(17.5,44.5){\makebox(5,1){\large\bf \dots}}
\put(17.5,39.5){\makebox(5,1){\large\bf \dots}}
\put(17.5,34.5){\makebox(5,1){\large\bf \dots}}

\put(25,45){\vector(1,0){5}} \put(25,45){\makebox(5,5){$a_i$}}
\put(25,40){\vector(1,0){5}} \put(25,40){\makebox(5,5){$b_j$}}
\multiput(25,35)(2,0){2}{\line(1,0){1}}
\put(29,35){\vector(1,0){1}} \put(24.5,35){\makebox(5,5){$0$}}

\put(30,30){\framebox(12.5,20){}}
\put(32.5,33.5){\dashbox(7.5,3){1}}
\put(32.5,38.5){\framebox(7.5,3){}}
\put(32.5,43.5){\framebox(7.5,3){}}
\put(41,48.5){\framebox(1.5,1.5){}}

\put(42.5,45){\vector(1,0){5}} \put(42.5,40){\vector(1,0){5}} 
\multiput(42.5,35)(2,0){2}{\line(1,0){1}}\put(46.5,35){\vector(1,0){1}} 

\put(47.5,30){\framebox(12.5,20){}}
\put(50,43.5){\framebox(7.5,3){}}
\put(50,38.5){\framebox(7.5,3){}}
\put(50,33.5){\dashbox(7.5,3){}}
\put(58.5,48.5){\framebox(1.5,1.5){}}

\put(60,45){\vector(1,0){5}} \put(60,40){\vector(1,0){5}} 
\multiput(60,35)(2,0){2}{\line(1,0){1}} \put(64,35){\vector(1,0){1}} 

\put(67.5,44.5){\makebox(5,1){\bf \dots}}
\put(67.5,39.5){\makebox(5,1){\bf \dots}}
\put(67.5,34.5){\makebox(5,1){\bf \dots}}

\put(75,45){\vector(1,0){5}} \put(75,40){\vector(1,0){5}} 
\multiput(75,35)(2,0){2}{\line(1,0){1}}
\put(79,35){\vector(1,0){1}} 

\put(80,30){\framebox(12.5,20){}}
\put(82.5,33.5){\dashbox(7.5,3){}}
\put(82.5,38.5){\framebox(7.5,3){}}
\put(82.5,43.5){\framebox(7.5,3){}}
\put(91,48.5){\framebox(1.5,1.5){}}

\put(92.5,45){\vector(1,0){5}} \put(92.5,40){\vector(1,0){5}} 
\multiput(92.5,35)(2,0){2}{\line(1,0){1}}
\put(96.5,35){\vector(1,0){1}} 

\put(52.5,54){\vector(-1,0){22.5}}
\put(52.5,51.5){\makebox(17.5,5){$\delta$ cells}}
\put(70,54){\vector(1,0){22.5}}

\put(45,0){\framebox(20,25){}}
\put(47.5,2.5){\dashbox(15,5){\tt start}}

\put(51,11){\framebox(8,3){\tt b}}
\put(51,16){\framebox(8,3){\tt a}}

\put(61,21){\framebox(4,4){\tt q}}

\put(31,15){\makebox(10,5){\tt ain}}
\put(40,17.5){\vector(1,0){5}}
\put(31,10){\makebox(10,5){\tt bin}}
\put(40,12.5){\vector(1,0){5}}
\put(27.5,2.5){\makebox(10,5){\tt startin}}

\multiput(40,5)(2,0){2}{\line(1,0){1}}  %
\put(44,5){\vector(1,0){1}}

\put(65,17.5){\vector(1,0){5}}
\put(70,15){\makebox(10,5){\tt aout}}
\put(65,12.5){\vector(1,0){5}}
\put(70,10){\makebox(10,5){\tt bout}}
\multiput(65,5)(2,0){2}{\line(1,0){1}}
\put(69,5){\vector(1,0){}}
\put(73,2.5){\makebox(10,5){\tt startout}}

\end{picture} 
\end{center}

\vspace*{3mm}

\begin{flushleft}
{\small\tt
if start then\\
$~~$begin\\
$~~$q := ain/bin;\\
$~~$aout := 0$~~~~${\bf\{}pad in zeros for vanishing terms in 
                                                \boldmath$\ol{A} \}$\\
$~~$end\\
else aout := ain - q*bin;\\
bout := b;$~~$b := bin;$~~~~${\bf\{}it takes 2 cycles for each b to pass a cell{\bf\}}\\
startout := start;$~~$start := startin.  
}
\end{flushleft}

\begin{center}
\begin{tabular}{ll}
\hspace*{-1em}
Figure~3: & Systolic subarray and its cell definition
           for realising a transformation $R_A$\hspace*{-0.8em}\\
\hline
\end{tabular}
\end{center}

It is easy to see that the leftmost cell performs $q := a_i/b_j$ in the 
first cycle and computes terms of $A$ in subsequent cycles.  The $q\,$s
computed by other cells are always zero, since terms of $A$ that have degree
higher than $i - \delta$ are zero.  The only function of these cells is
to shift the coefficients of $A$ faster than those of $B$ (notice that each
coefficient of $B$ stays in each cell for two cycles).  Thanks to these 
``shifting'' cells the nonzero leading term $\ol{a}_{i-\delta}$ of $\ol{A}$
will emerge from the rightmost cell at the same cycle as $b_j\,$, the 
nonzero leading term of $B\,$.  Thus $\ol{a}_{i-\delta}$ and $b_j$ are
aligned to enter another subarray of cells to the right in order to realise
whatever transformation follows $T\,$. \\

Note that there is no need to keep track of the value of $\delta$ in the
systolic subarray.  If $\ol{A}$ is nonzero, the realisation of the 
transformation following $T$ starts automatically at the first cell that 
sees a nonzero input (i.e. $\ol{a}_{i-\delta}$) on it input line (denoted by
{\tt ain}).  If $\ol{A}$ is the zero polynomial then $T$ must be the last
transformation $T_k\,$  In this case, the coefficients of $\ol{B}$ will
continue being shifted to the right to be output from the rightmost cell, and 
they will form terms in the desired GCD.

\pagebreak[4]
\subsubsection{A design using the difference of degrees}
\medskip

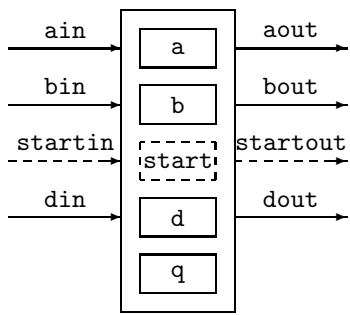
\begin{center}
\setlength{\unitlength}{1mm}
\begin{picture}(45,40)(0,0)
\small
\put(15,0){\framebox(15,40){}}
\put(17.5,32.5){\framebox(10,5){\tt a}}
\put(17.5,25){\framebox(10,5){\tt b}}
\put(17.5,17.5){\dashbox(10,5){\tt start}}
\put(17.5,10){\framebox(10,5){\tt d}}
\put(17.5,2.5){\framebox(10,5){\tt q}}

\put(0,35){\vector(1,0){15}} \put(0,35){\makebox(15,5){\tt ain}}
\put(0,27.5){\vector(1,0){15}} \put(0,27.5){\makebox(15,5){\tt bin}}
\multiput(0,20)(2,0){7}{\line(1,0){1}}
\put(13,20){\vector(1,0){2}}
\put(0,20){\makebox(15,5){\tt startin}}
\put(0,12.5){\vector(1,0){15}} \put(0,12.5){\makebox(15,5){\tt din}}

\put(30,35){\vector(1,0){15}} \put(30,35){\makebox(15,5){\tt aout}}
\put(30,27.5){\vector(1,0){15}} \put(30,27.5){\makebox(15,5){\tt bout}}
\multiput(30,20)(2,0){7}{\line(1,0){1}}
\put(43,20){\vector(1,0){2}}
\put(30,20){\makebox(15,5){\tt startout}}
\put(30,12.5){\vector(1,0){15}} \put(30,12.5){\makebox(15,5){\tt dout}}
\end{picture}
\end{center}

\begin{flushleft}
{\small\tt
dout := d;\\
startout := start;\\
$~~$case state  \{possible states are initial, reduceA and reduceB\} of\\
$~~$initial: \{wait for the beginning of a GCD computation\}\\
$~~~~$begin\\
$~~~~$aout := a;  bout := b;\\
$~~~~$if start then\\
$~~~~~~$begin\\
$~~~~~~$if (ain = 0) or ((bin $\neq$ 0) and (din $\geq$ 0)) then\\
$~~~~~~~~$begin\\
$~~~~~~~~$state := reduceA;\\
$~~~~~~~~$if bin = 0 then q := 0 else q := ain/bin; a := 0;\\
$~~~~~~~~$b := bin;  d := din $-$ 1\\
$~~~~~~~~$end\\
$~~~~~~$else\\
$~~~~~~~~$begin\\
$~~~~~~~~$state := reduceB;  q := bin/ain;  b := 0;\\
$~~~~~~~~$a := ain;  d := din + 1\\
$~~~~~~~~$end\\
$~~~~~~$end\\
$~~~~$end;\\
$~~$reduceA:  \{transform A and shift a's faster than b's\}\\
$~~~~$begin\\
$~~~~$if startin then state := initial;\\
$~~~~$aout := ain $-$ q*bin;  bout := b;  b := bin;\\
$~~~~$d := din\\
$~~~~$end;\\
$~~$reduceB:  \{transform B and shift b's faster than a's\}\\
$~~~~$begin\\
$~~~~$if startin then state := initial\\
$~~~~$aout := a;  a:= ain;  bout := bin $-$ q*ain;\\
$~~~~$d := din\\
$~~~~$end\\
$~~$end; \{case\}\\
start := startin.
}
\end{flushleft}

\begin{center}
\begin{tabular}{ll}
\hspace*{-0.5em}Figure~4: & Cell definition 
for the algorithm using differences of degrees\hspace*{-0.4em}\\
\hline
\end{tabular}
\end{center}

We have seen that a systolic subarray with cells defined as in Figure~3
can realise the transformation $T$ if it is known whether $T$ is $R_A$
or $R_B\,$.  Let $d = \deg A - \deg B\,$, where $A$ and $B$ are the
polynomials to be transformed by $T\,$.  Then $T$ is $R_A$ if $d \geq 0\,$,
otherwise $T$ is $R_B$ (see Section~2.1).  The cell definition of Figure~4
keeps track of the value of $d\,$, and consequently it is able to determine
on the fly which transformation to perform.  As in Figure~3, we specify the
cell using a Pascal-like language.  There are three states; operations
performed by each cell during a cycle depend on the state of the cell.
Initially, every cell is in state {\tt inital}.  Triggered by the {\tt start}
signal a cell will go to one of the other two states ({\tt reduceA} or
{\tt reduceB}) and eventually return to state {\tt initial}.\\

To illustrate the definition, consider once more the systolic subarray depicted
in Figure~3.  Suppose that $d = i - j > 0$ and $b_j \neq 0\,$.  Marching to the
right together with $b_j$ is the current value of $d\,$.  Each cell upon
receiving a true value on the systolic control part {\tt start} will go to
state {\tt reduceA} (since $d>0$).  When $\ol{a}_{i-\delta}$ $(\neq 0)$ and
$b_j$ are output from the rightmost cell of the subarray, they will enter
the cell to the right in the following cycle with state {\tt reduceA}
if $d\geq 0$ or {\tt reduceB} if $d<0$\,.\\

With $m+n+1$ cells a systolic array based on this design can compute a GCD of
any two polynomials of total degree less than $m+n+1\,$.  Moreover,
immediately after the input of one pair of polynomials, a new pair of
polynomials can enter the systolic array.  That is, the systolic array can
compute GCDs for multiple pairs of polynomials simutaneously, as they are
pipelined through the array.\\

We assume that none of the given pairs of polynomials have $x$ as a common
factor, so their GCDs have nonzero constant terms.  (A common power of $x$
can easily be factored out before the computation.)  With this assumption, one
can tell what the GCD is from the output emerging from the rightmost end of the 
array.  The constant term of the GCD is the last nonzero term emerging from the
array before output of the next batch of polynomials commences, and the  high 
degree terms of the GCD are those terms which emerged earlier on the
same output line.  If it is inconvenient to assume that the GCDs have nonzero
constant terms, one can either keep the degrees explicitly (instead of just
their difference) or have a ``stop'' bit to indicate the location of the 
leftmost of $a_0$ and $b_0\,$. 

\subsection{Some extensions}

The ``extended'' GCD problem is to find not only a greatest common divisor
of $G$ of $A_0$ and~$B_0\,$, but also polynomials $U$ and $V$ such that 
$UA_0 + VB_0 = G\,$.  The extended GCD problem is important for many 
applications,  including the decoder implementation for a variety of 
error-correcting codes.  The systolic array described above can be modified to 
compute $U$ and $V\,$:  see~[10] for details.\\

By keeping track of the beginning and end of each polynomial during the
computation, it is possible to avoid explicitly using the difference of
degrees of the polynomials (and hence no upper bound on this difference need
be known when the cells are designed).  Also, by interchanging $A$ and $B$ as
necessary, we can ensure that the output GCD always emerges on a fixed
output line.  These modifications lead to systolic algorithms whose 
implementations require fewer I/0 pins, which is an important practical
consideration.  The cell definition for one such algorithm is given in
Appendix A:  by interchanges it ensures that $d \geq 0\,$, and $d$ is
represented in unary as the distance between 1-bits on the {\tt start}
and {\tt sig} control paths.

\pagebreak[4]
\section{Integer GCD computation}

Consider now the problem of computing the greatest common divisor GCD {\tt (a, b)}
of two positive integers {\tt a} and {\tt b}, assumed to be given in the usual binary
representation.  Our aim is to compute GCD {\tt (a, b)} in time $O(n)$ on a linear
systolic array, where $n$ is the number of bits required to represent {\tt a} and {\tt b}
(say {\tt a < 2$^{\tt n}$, b < 2$^{\tt n}$}).  The significant difference between integer and
polynomial GCD computations is that carries have to be propagated in the former,
but not in the latter.\\

The classical Euclidean algorithm [32] may be written as:\\

{\small\tt\boldmath  while b $\neq$ 0  do
$\left\lgroup \begin{array}{c}
         {\tt a} \\  \\ {\tt b}
         \end{array} \right\rgroup$ := 
$\left\lgroup \begin{array}{c}
         {\tt b} \\  \\ {\tt a~mod~b}
         \end{array} \right\rgroup$ ; GCD := a .} \\

This is simple, but not attractive for a systolic implementation because the division
in the inner loop takes time $\Omega(n)$\,. More attractive is the ``binary''
Euclidean algorithm [8, 32] which uses only additions, shifts and comparisons.

\begin{flushleft}
{\small\tt
\{assume a, b odd for simplicity\}\\
t := |a - b|;\\
while t $\neq$ 0 do\\
$~~$begin\\
$~~$repeat t := t div 2 until odd(t);\\
$~~$if a > b then a := t else := t;\\
$~~$t := |a - b|\\
$~~$end;\\
GCD := a.}
\end{flushleft}

However, if we try to implement the binary Euclidean algorithm on a systolic array
we encounter a serious difficulty:  the test ``\,{\tt if  a $\geq$ b \dots}\,'' may require
knowledge of all bits of {\tt a} and {\tt b}, so again the inner loop takes time $\Omega(n)$
in the worst case.

\subsection{Algorithm PM}

\begin{flushleft}
{\small\tt\boldmath
\{assume a odd and b $\neq$ 0 , |a| $\leq$ 2$^{\tt n}$ , |b| $\leq$ 2$^{\tt n}$\}\\
$\alpha$ := n; $~\beta$ := n;\\
$~~$repeat\\
$~~$while even(b) do begin b := b div 2;  $~\beta$ := $\beta -$1 end;\\
$~~$\{now b odd, |b| $\leq$ 2$^{\tt \beta}$\}\\
$~~$if $\alpha \geq \beta$ then begin swap (a, b);  swap ($\alpha, \beta$) end;
                                         \{ "swap" has obvious meaning\}\\
$~~$\{now  $\alpha \leq \beta$  , |a| $\leq$ 2$^{\tt \alpha}$, |b| $\leq$ 2$^{\tt \beta}$,
                                                               a odd, b odd\}\\
$~~$if ((a+b) mod 4) = 0 then b := (a+b) div 2 else b := (a-b) div 2;\\
$~~$\{now b even, |b| $\leq$ 2$^{\tt \beta}$\}\\
$~~$until b = 0;\\
GCD := |a|.
}\end{flushleft}

\begin{center}
\ul{Figure~5:  Precursor to Algorithm PM}\\
\end{center}

Algorithm PM (for ``plus-minus''), like the classical and binary Euclidean algorithms,
finds the GCD of two $n$-bit integers {\tt a} and {\tt b} in $O(n)$ iterations, but we
shall see that it can be implemented in a pipelined fashion (least significant bits
first) on a systolic array.  Before defining Algorithm PM we consider the ``precursor''
algorithm defined in Figure~5.  Using the assertions contained in braces,
it is easy
to prove that the algorithm terminates in at most $2n + 1$ iterations (since $\alpha +
\beta$ strictly decreases at each iteration of the repeat block, except possibly the
first).\linebreak  
Moreover, since all transformations on {\tt a} and {\tt b} are GCD-preserving,
the GCD is computed\linebreak 
~\\[-14pt]
correctly.\\

It is not necessary to maintain $\alpha$ and $\beta$: all we need is their difference
$\delta = \alpha - \beta$ (analogous to the difference of degrees in Section~2).  This
observation leads to Algorithm PM, which is defined in Figure~6.

\begin{flushleft}
{\small\tt \boldmath
\{assume a odd, b $\neq$ 0\}\\
$\delta$ := 0;\\
$~~$repeat\\
$~~$while even(b) do begin b := b div 2; $\delta$ := $\delta$ + 1 end;\\
$~~$if $\delta \leq$ 0 then begin swap (a, b);  $~\delta$ := $-\delta$ end;\\
$~~$if ((a+b) mod 4) = 0 then b := (a+b) div 2 else b := (a-b) div 2\\
$~~$until b = 0;\\
GCD := |a|.
}\end{flushleft}

\begin{center}
\ul{Figure~6: Algorithm PM}
\end{center}

\subsection{Implementation on a systolic array}

For implementation on a systolic array, Algorithm PM has a great advantage over
the classical or binary Euclidean algorithms:  the tests in the inner loop involve
only the two least-significant bits of {\tt a} and {\tt b}\,.  Hence, a cell can
perform these tests before the high-order bits of {\tt a} and {\tt b} reach it via
cells to its left.  (The termination criterion ``\,{\tt until b = 0}\,'' is not a problem:
see below.)\\

We have to consider implementation of operations on $\delta$ in Algorithm PM.  The
only operations required are ``\,{\tt \boldmath $\delta := \delta + 1$}\,'',
``\,{\tt \boldmath $\delta := -\delta$}\,'',
and ``\,{\tt\boldmath if $\delta \geq 0$ \dots}\,''. Rather than represent $\delta$ in
binary, we choose a ``sign and magnitude unary'' representation, i.e.~keep sign
$(\delta)$ and $|\delta|$ separate, and represent $\varepsilon = |\delta|$ in unary
as the distance between 1-bits in two stream of bits.  With this representation all
required operations on $\delta$ can be pipelined.\\

\begin{center}
\setlength{\unitlength}{1mm}
\begin{picture}(95,50)(0,0)
\put(22.5,0){\framebox(45,50){}}

\put(0,44){\hbox{\tt ain}} \put(7.5,45){\vector(1,0){15}}
\put(0,36.5){\hbox{\tt bin}} \put(7.5,37.5){\vector(1,0){15}}
\put(0,29){\hbox{\tt startin}} \multiput(15,30)(2,0){3}{\line(1,0){1}}
                               \put(20,30){\vector(1,0){1}}
\put(0,21.5){\hbox{\tt startoddin}}
                                \put(20,22.5){\vector(1,0){2.5}}
\put(0,14){\hbox{\tt epsin}} \multiput(10.5,15)(2,0){4}{\line(1,0){1}}
                              \put(20,15){\vector(1,0){1}}
\put(0,6.5){\hbox{\tt negin}} \multiput(10.5,7.5)(2,0){4}{\line(1,0){1}}
                               \put(20,7.5){\vector(1,0){1}}

\put(25,42.5){\framebox(17,5){\tt a}}
\put(25,35){\framebox(17,5){\tt b}}
\put(25,27.5){\dashbox(17,5){\tt start}}
\put(25,20){\dashbox(17,5){\tt startodd}}
\put(25,12.5){\dashbox(17,5){\tt eps}}
\put(25,5){\dashbox(17,5){\tt neg}}

\put(48,42.5){\dashbox(17,5){\tt wait}}
\put(48,35){\dashbox(17,5){\tt shift}}
\put(48,27.5){\dashbox(17,5){\tt carry}}
\put(48,20){\dashbox(17,5){\tt swap}}
\put(48,12.5){\dashbox(17,5){\tt eps2}}
\put(48,5){\dashbox(17,5){\tt minus}}

\put(77.5,44){\hbox{\tt aout}} \put(67.5,45){\vector(1,0){7.5}}
\put(77.5,36.5){\hbox{\tt bout}} \put(67.5,37.5){\vector(1,0){7.5}}
\put(77.5,29){\hbox{\tt startout}} \multiput(67.5,30)(2,0){3}{\line(1,0){1}}
                                      \put(73.5,30){\vector(1,0){1}}
\put(77.5,21.5){\hbox{\tt startoddout}} \multiput(67.5,22.5)(2,0){3}{\line(1,0){1}}
                                     \put(73.5,22.5){\vector(1,0){1}}
\put(77.5,14){\hbox{\tt epsout}}\multiput(67.5,15)(2,0){3}{\line(1,0){1}}
                                     \put(73.5,15){\vector(1,0){1}}
\put(77.5,6.5){\hbox{\tt negout}}\multiput(67.5,7.5)(2,0){3}{\line(1,0){1}}
                                      \put(73.5,7.5){\vector(1,0){1}}

\end{picture}

\vspace*{3mm}
{\begin{tabular}{ll}
\hspace*{-0.8em}
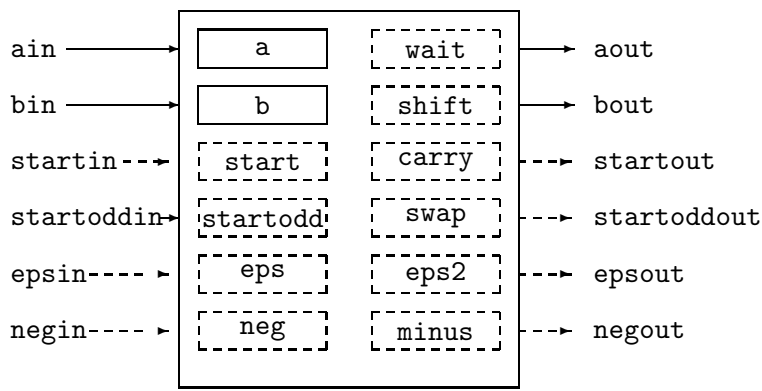
Figure~7: & Systolic cell for integer GCD computation\hspace*{-0.6em}\\
\hline
\end{tabular}}
\end{center}

After some optimisations we obtain the systolic cell illustrated in
Figure~7 and defined in\linebreak 
Appendix~B.
The cell has six input streams (each one bit wide): {\tt ain} and
{\tt bin} for the bits of the numbers {\tt a} and {\tt b} represented in 2's
complement binary (least significant bit first),
{\tt startin} to indicate the least
significant bit of {\tt a}, and three additional streams
{\tt startoddin}, {\tt epsin}
and {\tt negin} which should be all zero on input to the leftmost cell.
{\tt startoddin}
is used to indicate the least significant 1-bit of {\tt a} or {\tt b},
{\tt epsin}
and {\tt negin} are used to represent $\delta$\,.
There are six corresponding output
streams (connected, of course, to the input streams
of the cell to the right).  The cell has twelve internal state bits:  one for
each of the six inputs and six additional bits
({\tt wait, shift,}\linebreak
~\\[-15pt]
{\tt{carry, swap, eps2}} and {\tt minus}).\\

The termination criterion {\tt (b = 0)} need not be checked because once {\tt b} is
reduced to zero, cells further to the right will %
implement the statement
``\,{\tt\boldmath begin b := b div 2; $\delta$ := $\delta$ + 1 end}\,'' (see Figure~6)
and transmit {\tt a} unchanged, so the correct result will emerge from the rightmost
cell.  All we need is at least $4n$ cells to guarantee that Algorithm PM has reduced
{\tt b} to zero.  Actually, $3.1106n + 1$ cells suffice:  see [11].  Note that the
final output may represent $-$GCD ({\tt a,b}) in 2's complement: an additional
$O(n)$ cells are required to ensure that the output is +GCD ({\tt a,b}).\\

The definition of the cell illustrated in Figure~7 is given in Appendix B.  It
implements Algorithm PM (see Figure~6) with a straightforward modification to allow even
inputs as well as odd.

\pagebreak[3]
\section{Solution of Toeplitz systems}

A Toeplitz matrix $A = (a_{ij})$ is one which is constant along each diagonal, i.e.
$a_{lj}$ is a function of $j - i$ (which we denote by $a_{j-i}$).  
We are interested 
in the solution of a Toeplitz system of linear equations
$A\ul{x} = \ul{b}$\,,\\

\hspace*{30mm}\setlength{\unitlength}{1mm}
\begin{picture}(32,20)(-5,0)

\put(-36,8){where $A = (a_{j-i})$ =}
\put(34,9){,}

\put(0,15){\makebox(8,5){$a_0$}}
\put(0,10){\makebox(8,5){$a_{-1}$}}
\put(0,6){\makebox(8,5){$\vdots$}}
\put(0,0){\makebox(8,5){$a_{-n}$}}

\put(8,15){\makebox(8,5){$a_1$}}
\put(8,10){\makebox(8,5){$a_0$}}
\put(8,10){\line(3,-2){8}}
\put(8,0){\makebox(8,5){$\dots$}}

\put(16,15){\makebox(8,5){$\dots$}}
\put(16,15){\line(3,-2){8}}
\put(16,10){\line(3,-2){8}}
\put(16,0){\makebox(8,5){$a_{-1}$}}

\put(24,15){\makebox(8,5){$a_n$}}
\put(24,11){\makebox(8,5){$\vdots$}}
\put(24,5){\makebox(8,5){$a_1$}}
\put(24,0){\makebox(8,5){$a_0$}}

\put(0,0){\line(0,1){20}}
\put(0,0){\line(1,0){1}}
\put(0,20){\line(1,0){1}}

\put(32,0){\line(0,1){20}}
\put(32,0){\line(-1,0){1}}
\put(32,20){\line(-1,0){1}}

\end{picture}\\[-23mm]

\hspace*{74mm}
$\ul{b} =  \left[ \begin{array}{c}
                    b_0   \\
                    \vdots  \\
                    b_n\\
                    \end{array} \right] $\,, $\;\;$ and
$\ul{x} =  \left[ \begin{array}{c}
                    x_0   \\
                    \vdots  \\
                    x_n\\
                    \end{array} \right] $\,.\\[20pt]

(It is convenient to consider $(n+1)$-vectors and $(n+1)$ by $(n+1)$ Toeplitz matrices, 
with indices running from 0 to $n$\,.)  Large Toeplitz systems often arise in filtering 
and signal processing applications [1, 40, 56]: values of $n$ greater than 1000 are
common, so it is important to have special algorithms which take advantage of the
Toeplitz structure.  In applications $A$ is often symmetric positive-definite, but
we do not assume this here.\\

Several serial algorithms which require time $O(n^2)$ are known for the solution of
Toeplitz systems, for examples see [3, 31, 44, 58, 62].  Serial algorithms requiring
time $O(n \log^2n)$ and space $O(n)$ are also known [4, 9, 48], although their
practical value is not yet clear [54].

\subsection{Systolic algorithms for Toeplitz systems}

It is natural to ask if a linear systolic array of $O(n)$ cells can be used to solve
a Toeplitz system in time $O(n)$\,.  This is indeed the case [1, 41, 42, 49], but
the systolic algorithms presented in the cited papers have two shortcomings.\\

\begin{tabular}{ll}
1.& They assume that $A$ is symmetric, and\\
2.& they use $\Omega(n^2)$ storage, i.e.~$\Omega(n)$ words per cell.
\end{tabular}\\

We shall outline a systolic algorithm which avoids both these shortcomings: it 
applies to unsymmetric Toeplitz systems (although it may be specialised to the 
symmetric case with some savings if desired), and the total storage required is $O(n)$
words, i.e.~only a constant per cell.  (We consider words of storage rather than bits:
a word is assumed to be large enough to hold one floating-point number or an integer
of size $O(n)$\,, although the latter requirement could be avoided.)

\pagebreak[4]
\subsection{The algorithm of Bareiss}

Our systolic architecture uses an algorithm of Bareiss [3] to compute (implicitly)
an LU factorisation of the Toeplitz matrix $A$\,.  Historically the Bareiss recursions
in the symmetric case are due to Schur [53].\\

Define the ``shift'' matrix $Z_k$ by $Z_k = (z^{(k)}_{ij})$\,,\\

\hspace*{38.5mm}where $z^{(k)}_{ij} = \left\{ \begin{array}{l}
                      1 \hbox{ if } j-i = k\\
                       0 \hbox{ otherwise }\,.
                       \end{array} \right. $\\

Thus, premultiplication of $A$ by $Z_k$ shifts the rows of $A$ up $k$ places with zero
fill.  The Bareiss algorithm is defined in Figure~8.  At a typical stage of the Bareiss
algorithm, the matrices $A^{(-k)}$ and $A^{(+k)}$ have the structure illustrated in
Figure~9 $(k = 0, \dots , n)$\,.\\

The Bareiss algorithm computes the same LU factorisation of $A$ as would be obtained by
Gaussian elimination without pivoting:  $U = A^{(-n)}$ and $a_0L = (A^{(+n)})^{T2}$\,,
where the superscript ``$\,T2\,$'' denotes reflection in the main antidiagonal.  It is
assumed that all leading principal minors of $A$ are nonsingular, so the LU
factorisation of $A$ exists.  By transforming the right-hand side as shown in
Figure~8, we obtain an upper triangular system $A^{(-n)}\ul{x} = \ul{b}^{(-n)}$\,,
so $A^{(+n)}$ may be discarded if our objective is merely to solve $A\ul{x} = \ul{b}$\,.\\

Because $A^{(-n)}$ is not Toeplitz, the Bareiss algorithm appears to require
$\Omega(n^2)$ storage.  However, at the expense of some extra computation, we can
get by with $O(n)$ storage.  The key idea is that we can run the Bareiss algorithm
backwards to regenerate the elements of $A^{(-n)}$ as they are required to solve
the triangular system $A^{(-n)}\ul{x} = \ul{b}^{(-n)}$ by ``back-substitution'' [60],
using the Toeplitz structure of $\ul{\ul{\beta}}$ and $\ul{\ul{\delta}}$ (see
Figure~9) and the equations \\

$\begin{array}{llcl}
\hspace*{48mm}& A^{(k-1)} &=& A^{(k)} + m_k Z_k A^{(-k)}\\
\hbox{and}&           &&  \\
          & A^{(1-k)} &=& A^{(-k)} + m_{-k} A_{-k} A^{(k-1)}
\end{array}$\\
\bigskip

\begin{center}
\ul{~~~~~~~~~~~~~~~~~~~~~~~~~~~~~~~}\\
\end{center}
{\small\tt\boldmath
A$^{(0)}$ := A; \ul{b}$^{(0)}$ := \ul{b};\\[1ex]
for k := 1 to n do\\[1ex]
$~~$begin\\[1ex]
$~~$m$_{-k}$ := a$^{(1-k)}_{k,0}${\big /}a$_{0}$; \
                              \{diagonal elements of A$^{(k-1)}$ equal a$_0$\}\\[1ex]
$~~$A$^{(-k)}$ := A$^{(1-k)} ~-~ $m$_{-k}$Z$_{-k}$A$^{(k-1)}$;
               \{only store \ul{\ul{$\alpha$}} and \ul{\ul{$\beta$}} : see Figure~9\}\\[1ex]
$~~$\ul{b}$^{(-k)}$ := \ul{b}$^{(1-k)} ~-~ $m$_{-k}$Z$_{-k}$\ul{b}$^{(k-1)}$;\\[1ex]
$~~$m$_{+k}$ := a$^{(k-1)}_{0,k}${\big /} a$^{(-k)}_{n,n}$ ;\\[1ex]
$~~$A$^{(+k)}$ := A$^{(k-1)} ~-~ $m$_{+k}$Z$_{+k}$A$^{(-k)}$;
                  \{only store \ul{\ul{$\gamma$}} and \ul{\ul{$\delta$}} : see Figure~9\}\\[1ex]
$~~$\ul{b}$^{(+k)}$ := \ul{b}$^{(k-1)} ~-~ $m$_{+k}$Z$_{+k}$\ul{b}$^{(-k)}$\\[1ex]
$~~$end.\\[1ex]
\{now A$^{(-n)}$\ul{x} = \ul{b}$^{(-n)}$ is an upper triangular system\}
}\\

\begin{center}
\ul{Figure~8:  The Bareiss algorithm}\\
\end{center}

\pagebreak[4]
\begin{center}
\setlength{\unitlength}{1mm}
\begin{picture}(110,40)(0,0)

\put(2.5,12.5){\line(0,1){25}}
\put(2.5,12.5){\line(1,0){1}}
\put(2.5,37.5){\line(1,0){1}}
\put(27.5,12.5){\line(0,1){25}}
\put(27.5,12.5){\line(-1,0){1}}
\put(27.5,37.5){\line(-1,0){1}}

\put(5,15){\line(1,0){10}}
\put(5,15){\line(0,1){10}}
\put(5,25){\line(1,-1){10}} %
\put(5,22){\line(1,-1){7}}
\put(5,19){\line(1,-1){4}}
\put(25,35){\line(-1,0){20}}
\put(25,35){\line(0,-1){20}}
\put(5,35){\line(1,-1){20}} %
\put(15,25){\line(1,0){10}} %
\put(12.5,27.5){\line(1,0){12.5}}
\put(10,30){\line(1,0){15}}
\put(7.5,32.5){\line(1,0){17.5}}
\put(18,25){\line(1,-1){7}} %
\put(21,25){\line(1,-1){4}}

\put(30,29){$\Bigg\}$}
\put(30,19){$\Bigg\}$}
\put(32,30){\line(1,0){5}}
\put(32,20){\line(1,0){5}}
\put(40,29){$k$ rows upper triangular, not Toeplitz}
\put(40,19){$n+1-k$ rows upper triangular, Toeplitz $(\ul{\ul{\beta}})$ }

\put(5,9){\Large$\underbrace{\hspace*{10mm}}$}
\put(10,7){\line(0,-1){4}}
\put(10,3){\line(1,0){27}}
\put(40,2){$n-k$ rows lower triangular, Toeplitz $(\ul{\ul{\alpha}})$ }

\put(15,11){\Large$\underbrace{\hspace*{10mm}}$}
\put(20,9){\line(0,-1){1}}
\put(20,8){\line(1,0){17}}
\put(40,7){$k$ zero diagonals}

\put(7.5,17.5){\boldmath$\ul{\ul{\alpha}}$}
\put(21.5,21){\boldmath$\ul{\ul{\beta}}$}

\end{picture}\\

\noindent %
\hspace*{5mm}\setlength{\unitlength}{1mm}
\begin{picture}(110,40)(0,0)       

\put(2.5,12.5){\line(0,1){25}}
\put(2.5,12.5){\line(1,0){1}}
\put(2.5,37.5){\line(1,0){1}}
\put(27.5,12.5){\line(0,1){25}}
\put(27.5,12.5){\line(-1,0){1}}
\put(27.5,37.5){\line(-1,0){1}}
\put(5,15){\line(1,0){20}}
\put(5,15){\line(0,1){20}}
\put(5,35){\line(1,-1){20}} %
\put(5,32){\line(1,-1){7}} 
\put(5,29){\line(1,-1){3.75}} 
\put(5,25){\line(1,0){10}}
\put(5,22.5){\line(1,0){12.5}}
\put(5,20){\line(1,0){15}}
\put(5,17.5){\line(1,0){17.5}}

\put(25,35){\line(-1,0){10}}
\put(25,35){\line(0,-1){10}}
\put(15,35){\line(1,-1){10}} %
\put(21,35){\line(1,-1){4}}  
\put(18,35){\line(1,-1){7}}  
\put(22,30){\boldmath$\ul{\ul{\delta}}$}
\put(7,27.5){\boldmath$\ul{\ul{\gamma}}$}

\put(30,29){$\Bigg\}$}
\put(30,19){$\Bigg\}$}
\put(32,30){\line(1,0){5}}
\put(32,20){\line(1,0){5}}
\put(40,29){$n-k$ rows upper triangular, Toeplitz $(\ul{\ul{\delta}})$ }
\put(40,19){$k$ zero diagonals }

\put(5,11){\Large$\underbrace{\hspace*{20mm}}$}
\put(15,9){\line(0,-1){4}}
\put(15,5){\line(6,1){12}}
\put(15,5){\line(6,-1){12}}
\put(30,6.5){lower triangle, top $n+1-k$ rows Toeplitz $(\ul{\ul{\gamma}})$ }
\put(30,1.5){bottom $k$ rows, not Toeplitz}

\end{picture}\\[10pt]

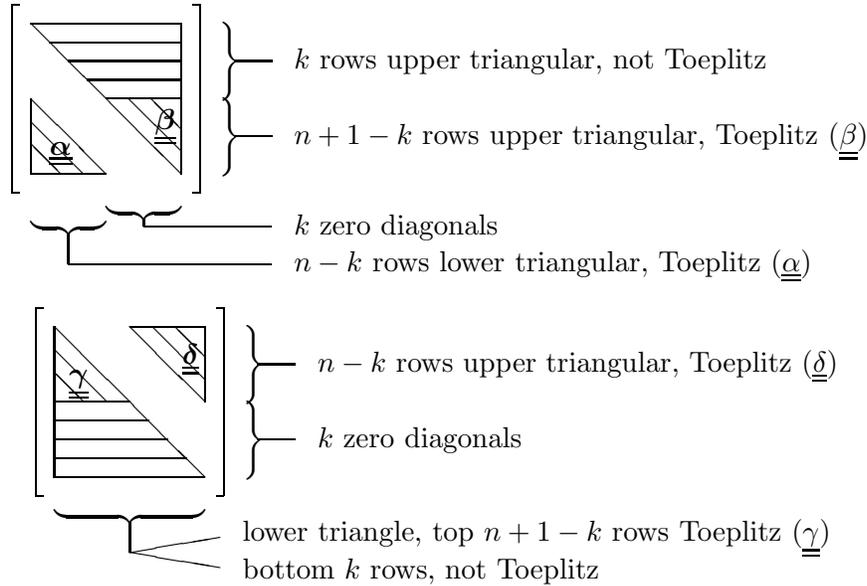
\ul{Figure~9:  Structure of $A^{(-k)}$ and $A^{(k)}$
in the Bareiss algorithm }
\end{center}

for $k=n, n-1, \dots , 1$\,.  (Observe that row $k$ of $A^{(-n)}$ is equal to row
$k$ of $A^{(-k)}$\,.)  Hence, only $O(n)$ storage is required to regenerate
rows $n, n-1, \dots , 0$ of $A^{(-n)}$\,: we need only to save the
multipliers $m_{\pm k}$ and the last column of $A^{(-n)}$\,.  In the systolic
algorithm described below these $O(n)$ numbers are simply saved in the appropriate
systolic cells.\\

A different way of reducing the storage requirements to $O(n)$ is to use the
Gohberg-Semencul formula [9] for the inverse of $A$\,, but the method outlined
above is simpler and can take advantage of any band structure in $A$ [15].

\subsection{A systolic algorithm for the solution of Toeplitz systems}

In the Bareiss algorithm four triangular Toeplitz matrices are updated (see
Figures 8 and 9).\\

\hspace*{40mm}\setlength{\unitlength}{1mm}
\begin{picture}(90,20)(-2,0)       

\put(1,8){$\ul{\ul{\alpha}}~=~ $}
\put(43,8){,}

\put(12,16){\makebox(6,4){$\alpha_0$}}
\put(18,16){\line(3,-2){18}}
\put(12,12){\makebox(6,4){$\alpha_1$}}
\put(18,12){\line(3,-2){12}}

\put(35,15){0}

\put(30,0){\makebox(6,4){$\alpha_1$}}
\put(36,0){\makebox(6,4){$\alpha_0$}}

\put(18,8){\line(3,-2){6}}

\put(12,0){\line(0,1){20}}
\put(12,0){\line(1,0){1}}
\put(12,20){\line(1,0){1}}

\put(42,0){\line(0,1){20}}
\put(42,0){\line(-1,0){1}}
\put(42,20){\line(-1,0){1}}

\put(51,8){$\ul{\ul{\beta}}~=~$}

\put(60,16){\makebox(6,4){$\beta_0$}}
\put(66,16){\line(3,-2){18}}
\put(66,16){\makebox(6,4){$\beta_1$}}
\put(72,16){\line(3,-2){12}}

\put(65,3){0}

\put(84,0){\makebox(6,4){$\beta_0$}}
\put(84,4){\makebox(6,4){$\beta_1$}}

\put(78,16){\line(3,-2){6}}

\put(60,0){\line(0,1){20}}
\put(60,0){\line(1,0){1}}
\put(60,20){\line(1,0){1}}

\put(90,0){\line(0,1){20}}
\put(90,0){\line(-1,0){1}}
\put(90,20){\line(-1,0){1}}

\end{picture}

\vspace*{3mm}

\hspace*{40mm}\setlength{\unitlength}{1mm} %
\begin{picture}(90,20)(-2,0)

\put(1,8){$\ul{\ul{\gamma}}~=~ $}
\put(43,8){,}

\put(12,16){\makebox(6,4){$\gamma_0$}}
\put(18,16){\line(3,-2){18}}
\put(12,12){\makebox(6,4){$\gamma_1$}}
\put(18,12){\line(3,-2){12}}

\put(35,15){0}

\put(30,0){\makebox(6,4){$\gamma_1$}}
\put(36,0){\makebox(6,4){$\gamma_0$}}

\put(18,8){\line(3,-2){6}}

\put(12,0){\line(0,1){20}}
\put(12,0){\line(1,0){1}}
\put(12,20){\line(1,0){1}}

\put(42,0){\line(0,1){20}}
\put(42,0){\line(-1,0){1}}
\put(42,20){\line(-1,0){1}}

\put(51,8){$\ul{\ul{\delta}}~=~$}

\put(60,16){\makebox(6,4){$\delta_0$}}
\put(66,16){\line(3,-2){18}}
\put(66,16){\makebox(6,4){$\delta_1$}}
\put(72,16){\line(3,-2){12}}

\put(65,3){0}

\put(84,0){\makebox(6,4){$\delta_0$}}
\put(84,4){\makebox(6,4){$\delta_1$}}

\put(78,16){\line(3,-2){6}}

\put(60,0){\line(0,1){20}}
\put(60,0){\line(1,0){1}}
\put(60,20){\line(1,0){1}}

\put(90,0){\line(0,1){20}}
\put(90,0){\line(-1,0){1}}
\put(90,20){\line(-1,0){1}}

\end{picture}\\[2ex]

We use a linear array of cells $P_0, P_1, \dots, P_n$ where cell $P_k$ has
registers to store $\alpha_k\,, \beta_k\,, \gamma_k$ and~$\delta_k$\,. (When describing
cell $P_k$ we omit the subscripts and simply refer to registers $\alpha, \beta,
\gamma$ and~$\delta$\,.)  Cell $P_k$ requires four additional registers:
$\lambda_k$ for a multiplier $m_{-j}$\,, $\mu_k$ for a multiplier $m_{+j}$\,,
and $\xi_k$ and $\eta_k$ which are associated with the right-hand side vector
$\ul{b}$ and the solution $\ul{x}$\,.

\pagebreak[4]
\begin{center}
Phase 1: LU decomposition by the Bareiss algorithm\\

{\small
\setlength{\unitlength}{1mm}
\begin{picture}(70,10)(0,0)       

\put(0,2.5){\framebox(7.5,5){\small$P_{k-1}$}}
\put(32.5,2.5){\framebox(5,5){\small$P_k$}}
\put(62.5,2.5){\framebox(7.5,5){\small$P_{k+1}$}}

\put(7.5,5.5){\makebox(25,5){$\lambda ~,~ \mu$}}
\put(7.5, 6){\vector(1,0){25}}
\put(37.5,5.5){\makebox(25,5){$\lambda  ~,~ \mu$}}
\put(37.5, 6){\vector(1,0){25}}

\put(7.5,-.5){\makebox(25,5){$\alpha ~,~ \delta ~,~ \xi $}}
\put(32.5,4){\vector(-1,0){25}}
\put(37.5,-.5){\makebox(25,5){$\alpha ~,~ \delta ~,~ \xi $}}
\put(62.5,4){\vector(-1,0){25}}

\end{picture} }\\[2ex]

Phase 2:  Back substitution to solve triangular system\\

{\small
\setlength{\unitlength}{1mm}
\begin{picture}(70,10)(0,0)       
\put(0,2.5){\framebox(7.5,5){\small$P_{k-1}$}}
\put(32.5,2.5){\framebox(5,5){\small$P_k$}}
\put(62.5,2.5){\framebox(7.5,5){\small$P_{k+1}$}}

\put(7.5,5.5){\makebox(25,5){$\delta ~,~ \xi$}}
\put(7.5, 6){\vector(1,0){25}}
\put(37.5,5.5){\makebox(25,5){$\delta ~,~ \xi$}}
\put(37.5, 6){\vector(1,0){25}}

\put(7.5,-.5){\makebox(25,5){$\lambda ~,~ \mu ~,~ \eta $}}
\put(32.5,4){\vector(-1,0){25}}
\put(37.5,-.5){\makebox(25,5){$\gamma ~,~ \mu ~,~ \eta $}}
\put(62.5,4){\vector(-1,0){25}}

\end{picture} }\\[2ex]

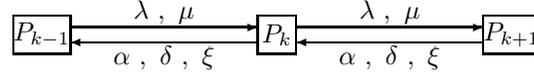
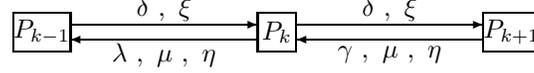
\ul{Figure~10:  Data flow for systolic Toeplitz solver}
\end{center}

Data flows in both directions between adjacent cells as shown in Figure~10.
Each cell needs five input and five output data paths, denoted by {\tt inL1,
inL2, inR1, inR2, inR3, outL1, outL2, outL3, outR1} and {\tt outR2} 
(see Figure~11).

\begin{center}
\setlength{\unitlength}{1mm}
\begin{picture}(92.5,40)(0,0)       

\put(20,0){\framebox(52.5,37.5){}}
\put(25,25){\framebox(5,5){$\alpha$}}
\put(37.5,25){\framebox(5,5){$\beta$}}
\put(50,25){\framebox(5,5){$\gamma$}}
\put(62.5,25){\framebox(5,5){$\delta$}}
\put(25,10){\framebox(5,5){$\lambda$}}
\put(37.5,10){\framebox(5,5){$\mu$}}
\put(50,10){\framebox(5,5){$\xi$}}
\put(62.5,10){\framebox(5,5){$\eta$}}

\put(0,31.5){\tt inL1} \put(10,32.5){\vector(1,0){10}}
\put(0,24){\tt inL2} \put(10,25){\vector(1,0){10}}
\put(0,16.5){\tt outL1} \put(20,17.5){\vector(-1,0){10}}
\put(0,9){\tt outL2} \put(20,10){\vector(-1,0){10}}
\put(0,1.5){\tt outL3} \put(20,2.5){\vector(-1,0){10}}

\put(83.5,31.5){\tt outR1} \put(72.5,32.5){\vector(1,0){10}}
\put(83.5,24){\tt outR2} \put(72.5,25){\vector(1,0){10}}
\put(83.5,16.5){\tt inR1} \put(82.5,17.5){\vector(-1,0){10}}
\put(83.5,9){\tt inR2} \put(82.5,10){\vector(-1,0){10}}
\put(83.5,1.5){\tt inR3} \put(82.5,2.5){\vector(-1,0){10}}
\end{picture}\\[2ex]

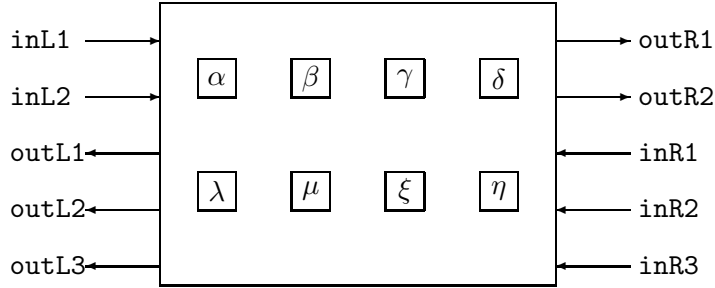
\ul{Figure~11:  Cell for systolic Toeplitz solver}
\end{center}

To avoid broadcasting multipliers $\lambda$ and $\mu$ during Phase 1, we use a
common technique [14, 37, 43]:  cells are active only on alternate time steps
$(P_0, P_2, \dots$ at time $T = 0, 2, \dots$ and $P_1, P_3 \dots$ at time
$T=1, 3, \dots)$\,, and the operation cell $P_k$ is delayed by $k$ time steps
relative to the operation of cell $P_0$\,.  A similar technique is used during
Phase 2, to avoid broadcasting $\delta$ and $\xi$\,.  (For another example of
this technique, see Section~5.3.)\\

\pagebreak[3]

Initialisation is as follows:\\

{\small\tt\boldmath
for k := 1 to n do\\
$~~$begin\\
$~~\alpha_k$ := a$_{-(k+1)}$; $\beta_k$ := a$_k$; $\gamma_k$ := a$_{-k}$;
                                     $\delta_k$ := a$_{k+1}$ ;\\
$~~\lambda_k$ := 0; $\mu_k$ := 0; $\xi_k$ := b$_{n-k-l}$; $\eta_k$ := b$_{n-k}$;\\
$~~$\{we assume that a$_{-(n+1)}$ = a$_{n+1}$ = b$_{-1}$ = 0 to cover end-conditions\}\\
$~~$end. }\\

Clearly this can be done in  time $O(n)$ if $A$ and $\ul{b}$ are available at
either end of the systolic array.\\

The definition of cell $P_k (0\leq k \leq n)$ is  given in Appendix C.  The final
solution $\ul{x}$ is given by $x_k = \xi_k$\,, where $\xi_k$ is stored in register
$\xi$ of processor $P_k$ after step $4n$\,.  We make some observations concerning the 
definition; for further details see [15]:

\begin{enumerate}
\item Cell $P_k$ is active only if $k \leq T < 2n-k$ (Phase 1) or $2n+k \leq T \leq
4n -k$ (Phase 2).  It is assumed that cell $P_k$ knows its index $k$ and the current
value of $T$ (though this could be avoided by the use of 1-bit systolic control paths
as in Section~3).
\item Pairs of adjacent cells could be combined, since only one cell of each pair
is active at each time step.  This would increase the mean cell utilisation from
25\% to 50\% (see observation 1 above).
\item Cell $P_0$ performs floating-point divisions, other cells perform only additions
and multiplications.  The total number of multiplications is $4.5n^2 + O(n)$\,.
A time step has to be long enough for six floating-point additions and multiplications,
plus data transfers, during Phase 1 (three during Phase 2); these may be performed
concurrently (with trivial modifications to the cell definition in Appendix C)
provided $P_0$ is sufficiently fast.
\item If $a_k = {\displaystyle\sum_j}~y_{j}y_{j+k}$ for some data $y_j$\,, as is common in
applications [40, 44], the coefficient $a_k$ can be computed in place by cell $P_k$\,.
\item Simplifications are possible in the symmetric case $(a_k = a_{-k})$\,.
\item The algorithm is numerically unstable in the general case, because it involves
the LU factorisation of $A$ without pivoting.  In fact, the algorithm breaks down if
a principal minor of $A$ is singular (e.g.~if $a_0 = 0$).  However, in applications
$A$ is often diagonally dominant or positive definite (see observation 4 above).  For
further discussion of the numerical properties of related algorithms, see [20, 56].
Sweet [56] gives an $O(n^2)$ time (serial) algorithm which computes an orthogonal
factorisation of $A$ and is numerically stable, but we do not know if it can be 
implemented in time $O(n)$ on a systolic array of $O(n)$ cells.
\item Cell $P_k$ typically reads its input lines  {\tt inL1,\dots,inR3}, does some
floating-point computations, and then writes to its output lines 
{\tt outL1,\dots,outR2}\,.  Hence, pairs of lines could be combined into single
bidirectional lines (e.g.~{\tt inL1} and {\tt outL1} could be combined).
\end{enumerate}

\pagebreak[3]
\section{The Symmetric Eigenvalue Problem}

In this section we consider the problem of computing the eigenvalues (and, if
required, the eigenvectors) of a real symmetric $n$ by $n$ matrix $A$\,, using
a systolic algorithm.  Unlike the problems considered above, this problem must
be solved by an iterative method.  Several authors [6, 28, 52] have suggested the
use of the QR algorithm, but this does not seem particularly well-suited to
parallel computation.  Instead, we resurrect the old-fashioned method of Jacobi [30],
since it is possible to implement it efficiently on a systolic array.  Sameh [51] 
suggested the use of Jacobi's method on a parallel computer; the idea of permuting
rows and columns (as directed in Section~5.2) to avoid global communication
requirements was first suggested by Brent and Luk~[14].\\

A ``sweep'' is defined in Section~5.1 below.  Suppose that the Jacobi method requires
$S$ sweeps for convergence to working accuracy.  For random symmetric matrices $A$
it is conjectured [16] that $S=O(\log n)$\,; in practice $S \leq 10$ for all
reasonable values of $n$ [14, 16, 50].  We sketch how a sweep can be performed in
time $O(n)$ on a square array of $[n/2]$ by $[n/2]$ systolic cells, so the symmetric
eigenproblem can be solved in time $O(nS)$\,.  The reader is referred to [13, 14] for
many details which are omitted here because of space limitations.

\pagebreak[4]
\subsection{The serial Jacobi method}

The Jacobi method generates a sequence $\{A_k\}$ of symmetric matrices by
$A_{k+1} := R^T_k A_k R_k$\,, where $R_k$ is a plane rotation and $A_1 = A$\,.
Let $R_k = \Big(r^{(k)}_{pq}\Big)$\,, $A_k = \Big(a^{(k)}_{pq}\Big)$\,, 
and suppose that 
$R_k$ represents a rotation in the $(i,j)$ plane, with $i <j$\,.  We have
\[\left[\begin{array}{ll}
r^{(k)}_{ii} &r^{(k)}_{ij}\\ 
r^{(k)}_{ji} &r^{(k)}_{jj}
\end{array}\right] \; = \;
\left[\begin{array}{rl}
\cos\theta & \sin\theta\\ 
-\sin\theta &\cos\theta\
\end{array}\right]\;,
\]
where the angle $\theta$ is chosen so as to reduce the $(i, j)$ element of $A_{k+1}$
to zero.  The formulae used to compute $\sin\theta$ and $\cos\theta$  are given by
Rutishauser [50].  The matrix $A_{k+1}$ differs from $A_k$ only in rows and columns
$i$ and $j$\,.\\

There remains the problem of choosing $(i,j)$\,, which is usually done according to
some fixed cycle.  It is desirable to go through all the off-diagonal elements
exactly once in any sequence (called a ``sweep'') of \hbox{$n(n-1)/2$} rotations.  A
simple sweep consists of the cyclic-by-rows ordering $(1,2), (1,3), \dots, (1,n),
(2,3),$ $ \dots, (2,n), (3,4), \dots, (n-1,n)$\,.  Forsythe and Henrici [23]
prove that the cyclic-by-rows Jacobi method always converges if
$\vert\theta\vert \leq \pi/4$ (which can always be enforced).  The serial Jacobi
method enjoys ultimate quadratic convergence [59].\\

Unfortunately, the cyclic-by-rows scheme is apparently not amenable to parallel
processing.  In Section~5.2 we represent an ordering which enables us to do
$\lfloor n/2 \rfloor$ 
rotations simultaneously.  The (theoretical) price is the loss of quaranteed
convergence.  Hansen [26] defines a ``preference factor'' when comparing different
orderings for the serial Jacobi method.  Our new ordering is in fact quite desirable,
even for serial computation, for it asymptotically optimises the preference factor
as $n \rightarrow \infty$\,.  Thus, although the convergence proof of [23] does not 
apply,  we expect convergence in practice to be faster than for the cyclic-by-rows
ordering, and simulation results [14, 16] support this conclusion.  To ensure
convergence we may adopt a ``threshold'' approach [60]:  associate a threshold value
with each sweep, and when making the transformation of that sweep, omit any rotation
if the doomed off-diagonal element is smaller in magnitude than the threshold.

\subsection{A semi-systolic algorithm for the symmetric eigenvalue problem}

We first describe a semi-systolic algorithm for implementing the Jacobi method.  The
algorithm is semi-systolic rather than systolic because it assumes the ability to
broadcast row and column rotation parameters (i.e.~$\sin\theta, \cos\theta$ values).
In Section~5.3 we show how to avoid broadcasting.\\

Assume for simplicity that $n$ is even.  We use a square array of $n/2$ by $n/2$
systolic cells, each cell  containing a 2 by 2 submatrix of $A$\,.  Initially cell
$P_{ij}$ contains

$$\left[ \begin{array}{ll}
a_{2i-1,2j-1} & a_{2i-1,2j} \\[2ex]
a_{2i,2j-1}   & a_{2i,2j}
          \end{array} \right] $$

for $i, j=1, \dots , n/2$\,, and $P_{ij}$ is connected to its nearest neighbours 
$P_{i\pm 1,j}$ and $P_{i,j\pm 1}$\,.\linebreak  
In general $P_{ij}$ contains four real
numbers 
{\small
$\left[\begin{array}{ll}
               \alpha_{ij}&\beta_{ij}\\ \gamma_{ij}&\delta_{ij}
                \end{array}\right]$\,,
} %
where $\alpha_{ij}=\alpha_{ji}$\,,
$\delta_{ij}=\delta_{ji}$ and $\beta_{ij} = \gamma_{ji}$ 
by symmetry of $A$\,.

\pagebreak[4]
The diagonal cells $P_{ii}$ $(i = 1, \dots , n/2)$ act differently 
from the off-diagonal cells 
$P_{ij}\linebreak 
(1 \leq i, j \leq n/2,$ $i \neq j)$.  At each time step the diagonal cell
$P_{ii}$ computes a rotation
{\small 
$\left[\begin{array}{rl} 
c_{i} & s_{i}\\
-s_{i} & c_{i}
\end{array}\right]$ 
}
to annihilate its off-diagonal
elements $\beta_{ii}$ and~$\gamma_{ii}$ (actually $\beta_{ii}=\gamma_{ii}$) and
update its diagonal elements $\alpha_{ii}$ and~$\delta_{ii}$ accordingly.  To complete
these rotations, the off-diagonal cells $P_{ij}$ $(i \neq j)$ must perform the
transformations

\[
\left[\begin{array}{ll}
\alpha_{ij} & \beta_{ij}\\
\gamma_{ij} & \delta_{ij}
\end{array}\right] \;:=\;
\left[\begin{array}{lr}
c_{i} & -s_{i}\\
s_{i} & c_{i}
\end{array}\right]\;
\left[\begin{array}{ll}
\alpha_{ij} & \beta_{ij}\\
\gamma_{ij} & \delta_{ij}
\end{array}\right]\;
\left[\begin{array}{rl}
c_{j} & s_{j}\\
-s_{j} & c_{j}
\end{array}\right]\,.
\]

We asume that the diagonal cell $P_{ii}$ broadcasts the rotation parameters $c_i$ and
$s_i$ to cells $P_{ij}$ and $P_{ji}$ $(j=1, \dots , n/2)$ 
in the same row and column.\\

To complete a step, columns and corresponding rows are interchanged between adjacent
cells so that a new set of $n$ off-diagonal elements is ready to be annihilated by the
diagonal cells during the next time step.  The interchanges are done in two sub-steps.
First, adjacent columns are interchanged according to the permutation
$$ P = (3 5 7 \dots (2n-1) (2n) (2n-2) (2n-4) \dots 4 2)\,.$$

Note that this is {\bf not} the ``perfect shuffle'' permutation; it is a permutation
used in the singular value decomposition algorithm of [13], and only requires 
nearest-neighbour communication between systolic processors.  Next, the same
permutation $P$ is applied to the rows, to maintain symmetry.  {From} Section~3
of [13] it is clear that a complete sweep is performed every $n-1$ steps, because 
each off-diagonal element of $A$ is moved into one of the diagonal cells in
exactly one of the $n-1$ steps.  This is illustrated for the case $n=8$ in
Figure~12.

\begin{center}
\setlength{\unitlength}{1mm}
\begin{picture}(90,75)(0,0)       
\put(20,67){\makebox(10,5){\ul{$P_{11}$}}}
\put(40,67){\makebox(10,5){\ul{$P_{22}$}}}
\put(60,67){\makebox(10,5){\ul{$P_{33}$}}}
\put(80,67){\makebox(10,5){\ul{$P_{44}$}}}
\put(0,58){\makebox(10,10){\tt \ul{step 0}}}
\put(0,48){\makebox(10,10){\tt \ul{step 1}}}
\put(0,38){\makebox(10,10){\tt \ul{step 2}}}
\put(0,28){\makebox(10,10){\tt \ul{step 3}}}
\put(0,18){\makebox(10,10){\tt \ul{step 4}}}
\put(0,8){\makebox(10,10){\tt \ul{step 5}}}
\put(0,-2){\makebox(10,10){\tt \ul{step 6}}}
\put(20,60){\framebox(10,5){$1~~~~2$}}
\put(20,50){\framebox(10,5){$1~~~~4$}}
\put(20,40){\framebox(10,5){$1~~~~6$}}
\put(20,30){\framebox(10,5){$1~~~~8$}}
\put(20,20){\framebox(10,5){$1~~~~7$}}
\put(20,10){\framebox(10,5){$1~~~~5$}}
\put(20,0) {\framebox(10,5){$1~~~~3$}}

\put(40,60){\framebox(10,5){$3~~~~4$}}
\put(40,50){\framebox(10,5){$2~~~~6$}}
\put(40,40){\framebox(10,5){$4~~~~8$}}
\put(40,30){\framebox(10,5){$6~~~~7$}}
\put(40,20){\framebox(10,5){$8~~~~5$}}
\put(40,10){\framebox(10,5){$7~~~~3$}}
\put(40,0) {\framebox(10,5){$5~~~~2$}}

\put(60,60){\framebox(10,5){$5~~~~6$}}
\put(60,50){\framebox(10,5){$3~~~~8$}}
\put(60,40){\framebox(10,5){$2~~~~7$}}
\put(60,30){\framebox(10,5){$4~~~~5$}}
\put(60,20){\framebox(10,5){$6~~~~3$}}
\put(60,10){\framebox(10,5){$8~~~~2$}}
\put(60,0) {\framebox(10,5){$7~~~~4$}}

\put(80,60){\framebox(10,5){$7~~~~8$}}
\put(80,50){\framebox(10,5){$5~~~~7$}}
\put(80,40){\framebox(10,5){$3~~~~5$}}
\put(80,30){\framebox(10,5){$2~~~~3$}}
\put(80,20){\framebox(10,5){$4~~~~2$}}
\put(80,10){\framebox(10,5){$6~~~~4$}}
\put(80,0) {\framebox(10,5){$8~~~~6$}}

\multiput(22,56)(0,2){2}{\line(0,1){1}}
\multiput(22,46)(0,2){2}{\line(0,1){1}}
\multiput(22,36)(0,2){2}{\line(0,1){1}}
\multiput(22,26)(0,2){2}{\line(0,1){1}}
\multiput(22,16)(0,2){2}{\line(0,1){1}}
\multiput(22,6)(0,2){2}{\line(0,1){1}}
\dashline{1}(30,55)(40,60)
\dashline{1}(30,45)(40,50)
\dashline{1}(30,35)(40,40)
\dashline{1}(30,25)(40,30)
\dashline{1}(30,15)(40,20)
\dashline{1}(30, 5)(40,10)
\dashline{1}(50,55)(60,60)
\dashline{1}(50,45)(60,50)
\dashline{1}(50,35)(60,40)
\dashline{1}(50,25)(60,30)
\dashline{1}(50,15)(60,20)
\dashline{1}(50, 5)(60,10)
\dashline{1}(70,55)(80,60)
\dashline{1}(70,45)(80,50)
\dashline{1}(70,35)(80,40)
\dashline{1}(70,25)(80,30)
\dashline{1}(70,15)(80,20)
\dashline{1}(70, 5)(80,10)
\dashline{1}(80,55)(70,60)
\dashline{1}(80,45)(70,50)
\dashline{1}(80,35)(70,40)
\dashline{1}(80,25)(70,30)
\dashline{1}(80,15)(70,20)
\dashline{1}(80, 5)(70,10)
\dashline{1}(60,55)(50,60)
\dashline{1}(60,45)(50,50)
\dashline{1}(60,35)(50,40)
\dashline{1}(60,25)(50,30)
\dashline{1}(60,15)(50,20)
\dashline{1}(60, 5)(50,10)
\dashline{1}(40,55)(30,60)
\dashline{1}(40,45)(30,50)
\dashline{1}(40,35)(30,40)
\dashline{1}(40,25)(30,30)
\dashline{1}(40,15)(30,20)
\dashline{1}(40, 5)(30,10)

\end{picture} 

\vspace*{3mm}
\begin{tabular}{ll}
\hspace*{-1em}
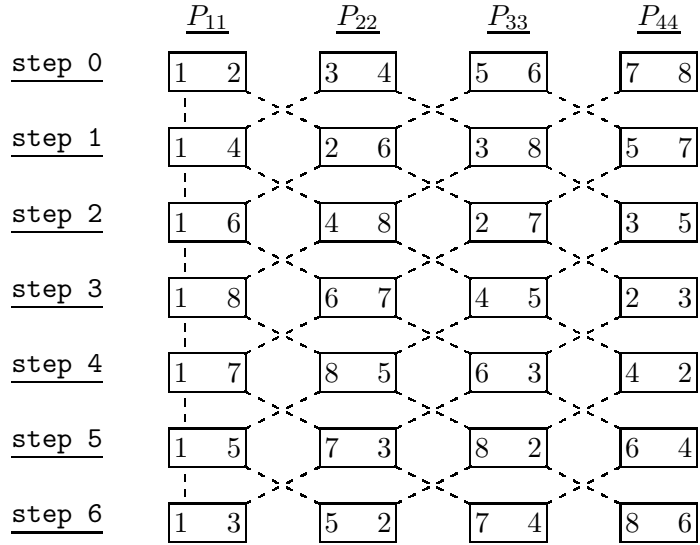
Figure~12: & Indices of off-diagonal elements in diagonal cells 
over a full sweep $(n=8)$\hspace*{-0.8em}\\
\hline
\end{tabular}
\end{center}

\pagebreak[4]
\subsection{Further details -- avoiding broadcast of rotation parameters}

In [14] details such as the threshold strategy, computation of eigenvectors, use of
diagonal connections between cells, taking full advantage of symmetry, handling the
case of odd $n$\,, etc are discussed.  Here we omit these details, but outline an
important point:  how to avoid broadcast of rotation parameters, i.e.~to convert the
semi-systolic algorithm of Section~5.2 into a systolic algorithm, while retaining
total running time $O(nS)$ for the algorithm.\\

Let $\Delta_{ij} = \vert i-j \vert$ denote the distance of cell $P_{ij}$ from the
diagonal.  The operation of cell $P_{ij}$ will be delayed by $\Delta_{ij}$ time units
relative to the operation of the diagonal cells, in order to allow time for rotation 
parameters to be propagated at unit speed along each row and column of the systolic
array.\\

A cell cannot commence the computations associated with a rotation until data from
earlier rotations is available on its input lines.  In particular, cell $P_{ij}$
needs data from cells $P_{i-1,j-1}, P_{i-1,j+1}, P_{i+1,j-1}$ and 
$P_{i+1,j+1}$
for $1 < i < n/2, 1 < j < n/2$ (the boundary cases are slightly different).  Since
$$ \vert \Delta_{ij} - \Delta_{i+1,j+1} \vert \leq 2 \,,$$

it is sufficient for cell $P_{ij}$ to be idle for two time steps while waiting for 
the cells $P_{i\pm 1, j\pm 1}$ to complete their (possibly delayed) steps.  Thus, the
price paid to avoid broadcasting rotation parameters is that each cell is active
for only one third of the total computation time.  A similar inefficiency occurs in
many other systolic algorithms, see for example [6, 12, 35, 37, 43] and Section~4.3.
In a practical design triples of three adjacent cells could share a floating-pont unit
to ameliorate this inefficiency.  Alternatively, ``idle'' cells could be used to 
increase the reliability of the systolic array by performing redundant computations
[33].

\subsection{Some extensions}

We have sketched how the symmetric eigenvalue problem can be solved in time $O(nS)$\,,
where $S$ is for practical purposes bounded by 10, using a square array of $O(n^2)$
systolic processors.  The speedup over the usual $O(n^3)$ serial algorithms
(e.g.~tridiagonalisation followed by the QR algorithm) is significant for moderate or 
large $n$\,.  Related algorithms for computing the singular value decomposition on a 
systolic array are presented in [13, 17].  For the unsymmetric eigenvalue problem
the question is open -- the ideas used in the symmetric case do not all carry over to
Eberlein's methods [22, 51] in an obvious way.  However, everything does carry over
with the obvious changes to complex Hermitian or normal matrices.

\section{Conclusion}

Systolic arrays provide cost-effective solutions to many important compute-bound
problems, although they are not a universal panacea.  The examples presented in
Sections 2--5 illustrate that the best serial algorithm does not always lead to the
best systolic algorithm.  A systolic array with $n$ cells can simulate (in real time)
a Turing machine which uses at most $n$ squares of tape, but a ``good'' systolic 
algorithm should be significantly faster than a simulation of a Turing machine.
There are many problems for which the existence of a good systolic algorithm remains
an open question.  Other open questions are:  how to compile code for a programmable
systolic array [21], how to prove the correctness of cell definitions such as those given
in Appendices A--C, and how best to implement the systolic cells. 
For example, should they 
use the bit-serial approach advocated in [1, 45, 47] or the bit-parallel approach of
[21]~?

\pagebreak[4]
\section*{Appendix A: Cell definition for systolic polynomial GCD\\
\hspace*{6em}computation}

{\small\tt\boldmath

\{The language used here and below is Pascal with some trivial extensions.  To save
space, \\ obvious declarations have been omitted.\}\\

aout := a;~~~~~~~~~a := ain;~~~~~~~~~\{standard transfers\}\\
bout := b;~~~~~~~~~b := bin;~~~~~~~~~\{assume deg B $\leq$ deg A\}\\
startout := start;~start := startin;~\{true for start of polynomial A \}\\
stopout := stop;~~~stop := stopin;~~~\{true for end of polynomial A \}\\
sigout := sig;~~~~~sig := sigin;~~~~~\{initially sig true if corresponding b $\neq$ 0\}\\
$~~~$case state \{possible states are initial, shift, swap and trans\} of\\
$~~~$initial:~~~\{wait for next start signal\}\\
$~~~~~~$if start and not stop then\\
$~~~~~~~~~$if b = 0 then state := shift else\\
$~~~~~~~~~~~~$begin q := a/b;~~\{division can be avoided\}\\
$~~~~~~~~~~~~$if sig then\\
$~~~~~~~~~~~~~~~$begin state := swap;~~a := b;~~sig := false end\\
$~~~~~~~~~~~~$else state := trans\\
$~~~~~~~~~~~~$end;\\
$~~~$shift:~~~\{shift~~B~~faster than~~A \}\\
$~~~~~~$begin bout := b;~~b := 0;\\
$~~~~~~$if stop then state := initial\\
$~~~~~~$end;\\
$~~~$swap:~~~\{transform, shift and interchange\}\\
$~~~~~~$begin bout := a - q*b;~~a := b;~~b := 0;\\
$~~~~~~$sig := (bout $\neq$ 0);\\
$~~~~~~$if stop then state := initial\\
$~~~~~~$end;\\
$~~~$trans:~~~\{transform, shift\}\\
$~~~~~~$begin aout := a - q*b;~~a := 0;\\
$~~~~~~$if stop then state := initial;\\
$~~~~~~$stopout := stop; stop := false;\\
$~~~~~~$sigout := sig; sig := false\\
$~~~~~~$end\\
$~~~$end \{case\}.
}

\pagebreak[4]

\section*{Appendix B: Cell definition for systolic integer GCD computation}

{\small\tt\boldmath

\{See Figure~7 for I/0 ports\}\\

aout := a;~~~~~~~~~~~~~~~~~~a := ain;~~~~\{standard transfers\}\\
bout := b;~~~~~~~~~~~~~~~~~~b := bin;\\
startout := start;~~~~~~~~~~start := startin;\\
startoddout := startodd;~~~~startodd := startoddin;\\
epsout := eps2; eps2 := eps;  eps := epsin; \{delay here\}\\
negout := neg;\\

wait := (wait or start) and not startodd; \{wait for nonzero bit\}\\

if startodd or (wait and (a or b)) then\\
$~~~$begin\\
$~~~$eps := eps or wait;\\
$~~~$eps2 := 0; \{0 $\equiv$ false,~~1 $\equiv$ true\}\\
$~~~$neg := negin and not wait;\\
$~~~$startodd := 1;\\
$~~~$wait := 0; \{end of waiting for a nonzero bit\}\\
$~~~$swap := not a;\\
$~~~$shift := not (a and b)\\
$~~~$end\\
else if wait then epsout := eps2 \{normal speed\}\\
else if shift then \{shift b faster than a, may also swap\}\\
$~~~$begin\\
$~~~$aout := (bout and swap)~~or~~(aout and not swap); \{normal speed\}\\
$~~~$bout := (a and swap)~~or~~(b and not swap);~~~~~~~\{fast speed\}\\
$~~~$epsout := (eps and neg)~~or~~(epsout and not neg);\\
$~~~$neg := neg and not (eps and startoddout); \{$\delta$ may become zero\}\\
$~~~$negout := neg\\
$~~~$end\\
else if startoddout then\\
$~~~$begin\\
$~~~$epsout := eps2;~~~~~~~~~~~~~~~~~~~~~\{normal speed\}\\
$~~~$swap := not neg;\\
$~~~$neg := neg or not eps2;~~~~~~~~~~~~~\{$\delta$ := -$|\delta|$\}\\
$~~~$negout := neg;\\
$~~~$aout := aout or swap;~~~~~~~~~~~~~~~\{swap implies b\}\\
$~~~$bout := 0;~~~~~~~~~~~~~~~~~~~~~~~~~~\{and new~~b~~is even\}\\
$~~~$carry := a $\oplus$ b;~~~~~~~~~~~~~~~~\{may be borrow or carry;  $\oplus$ is exclusive or\}\\
$~~~$minus := not carry~~~~~~~~~~~~~~\{1 iff we form (b - a) div 2\}\\
$~~~$end\\
else \{not startoddout\}\\
$~~~$begin\\
$~~~$epsout := eps2;~~~~~~~~~~~~~~\{normal speed\}\\
$~~~$aout := (bout and swap) or~~~(aout and not swap); \{normal speed\}\\
$~~~$bout := a $\oplus$ b $\oplus$ carry;~~~~~~\{fast speed\}\\
$~~~$carry := majority (b, carry, a$\;\oplus\;$minus) 
		\{majority true if 2 or 3 of its arguments true\}\\
$~~~$end.
}

\section*{Appendix C: Cell definition for systolic Toeplitz equation solver}

{\small\tt\boldmath

\{Program for cell k at time step T, 0 $\leq$ k $\leq$ n, 0 $\leq$ T $\leq$ 4n\}\\
\{See Figure~11 for I/0 ports\}\\

if even(T+k) and (T $\geq$ k) and (T<2n-k) then \{Phase 1 - LU factorisation\}\\
$~~~$begin\\
$~~~$if T > k then \{accept inputs from cell k+1\}\\
$~~~~~~$begin~~$\alpha$ := inR1;~~$\delta$ := inR2;~~$\xi$ := inR3 end;\\
$~~~$if k = 0 then \{compute multiplier\}~~$\lambda$ := $\alpha$/$\gamma$ else\\
$~~~~~~$begin \{accept multipliers from cell k-1\}\\
$~~~~~~$$\lambda$~~:= inL1;~~$\mu$  := inL2;\\
$~~~~~~$$\alpha$~~:= $\alpha$ - $\lambda$ * $\gamma$\\
$~~~~~~$end;\\
$~~~$$\beta$ := $\beta$ - $\lambda$ * $\delta$;~~$\eta$ := $\eta$ - $\lambda$ * $\xi$ ;\\ 
$~~~$if k = 0 then \{compute multiplier\}~~$\mu$ := $\delta$ / $\beta$~~else\\
$~~~~~~$begin\\
$~~~~~~$$\gamma$ := $\gamma$ - $\mu$ * $\alpha$ ;\\
$~~~~~~$$\delta$ := $\delta$ - $\mu$ * $\beta$ ;\\
$~~~~~~$$\xi$ := $\xi$ - $\mu$ * $\eta$\\
$~~~~~~$end;\\
$~~~$outL1 := $\alpha$ ;~~outL2 := $\delta$ ;~~outL3 := $\xi$ ; \{ignore outL1-3 if k = 0\}\\
$~~~$outR1 := $\lambda$ ;~~outR2 := $\mu$ ~~~~~~~~~~~~~~~~\{ignore outR1-2 if k = n\}\\
$~~~$end\\

else if even(T + k) and (T $\geq$ 2n+k) and (T $\leq$ 4n-k) then \{Phase 2 - back
                                                                      substitution\}\\
$~~~$begin\\
$~~~$if T > 2n+k then begin~~$\lambda$ := inR1;~~$\mu$ := inR2;~~$\eta$ := inR3 end;\\
$~~~$if k = 0 then begin~~$\xi$ := $\eta$ / $\beta$;~~$\delta$ := $\mu$ * $\beta$~~end else\\
$~~~~~~$begin\\
$~~~~~~$$\xi$ := inL1;~~$\delta$ := inL2;\\
$~~~~~~$$\eta$ := $\eta$ - $\beta$ * $\xi$ ; $\delta$ := $\delta$ + $\mu$ * $\beta$\\
$~~~~~~$end;\\
$~~~$$\beta$ := $\beta$ + $\lambda$ * $\delta$ ;\\
$~~~$outL1 := $\lambda$ ;~~outL2 := $\mu$ ;~~outL3 := $\eta$ ;~~\{ignore if k = 0\}\\  
$~~~$outR1 := $\xi$ ;~~outR2 := $\delta$ ~~~~~~~~~~~~~~~~~~\{ignore if k = n\}\\
$~~~$end.  
}

\pagebreak[3]
\subsection*{Acknowledgements}

{\small
The work of the first author was supported in part by the Australian Research Grants 
Scheme and in part by the Centre for Mathematical Analysis at the Australian National
University.  The work of the second author was supported in part by the Office of
Naval Research under contracts N00014-76-0270, NR 044-422 and N00014-80-C-0236,
NR 048-659, and in part by the Mathematical Sciences Research Centre, Australian
National University.  The work of the third author was supported in part by the
Mathematical Sciences Research Centre and the Centre for Mathematical Analysis,
Australian National University.
} %

\pagebreak[4]
\section*{References}

\begin{tabular}{rp{147mm}}
[1]& H.M.~Ahmed, J-M.~Delosme and M.~Morf,
Highly concurrent computing structures for matrix arithmetic and signal processing,
{\em IEEE Computer\/} 15, 1 (January 1982), 65--82.\\

$[2]$& A.V.~Aho, J.E.~Hopcroft and J.D.~Ullman,
{\em The Design and Analysis of Computer\linebreak 
Algorithms\/},
Addison-Wesley, Reading, Mass., 1974.\\

$[3]$& E.H.~Bareiss,
Numerical solution of linear equations with Toeplitz and vector Toeplitz matrices,
{\em Numer.~Math.\/} 13 (1969), 404--424.\\

$[4]$& R.R.~Bitmead and B.D.O.~Anderson,
Asymptotically fast solution of Toeplitz and related systems of linear equations,
{\em Linear Algebra and its Applications\/}
34 (1980), 130--116.\\

$[5]$& J.~Blackmer, P.~Kuekes and G.~Frank,
A 200 MOPS systolic processor,
{\em Proceedings of SPIE, Vol.~298:~Real-Time Signal Processing IV\/},
Society of Photo-Optical Instrumentation Engineers, Bellingham, Washington, 1981.\\

$[6]$&A.~Bojanczyk, R.P.~Brent and H.T.~Kung,
Numerically stable solution of dense systems of linear equations using
mesh-connected processors, to appear.
{[}Appeared in
{\em SIAM J.~Sci.~and Stat.~Comput.\/} 5 (1984), 95--104.{]}\\ 

$[7]$& A.~Borodin, J.~von zur Gathen and J.~Hopcroft,
Fast parallel matrix and GCD computations,
{\em Proceedings of the 23rd Annual Symposium on Foundations of Computer Science\/},
IEEE, New York, 1982, 65--71.\\

$[8]$& R.P.~Brent,
Analysis of the binary Euclidean algorithm,
in {\em New Directions and Recent Results in Algorithms and Complexity\/},
(J.F.~Traub, editor), Academic Press, New York, 1976, 321--355.\\

$[9]$& R.P.~Brent, F.G.~Gustavson and D.Y.Y.~Yun,
Fast solution of Toeplitz systems of equations and
computation of Pad\'e approximants,
{\em J.~Algorithms\/} 1 (1980), 259--295.\\

$[10]$& R.P.~Brent and H.T.~Kung,
{\em Systolic VLSI arrays for polynomial GCD computation\/},
Report CMU-CS-82-118, Department of Computer Science, Carnegie-Mellon University,
Pittsburgh, March 1982.
{[}Appeared in {\em IEEE Trans.\ on Computers} C--33 (1984),\linebreak 
731--736.{]}\\

$[11]$& R.P.~Brent and H.T.~Kung,
{\em A systolic VLSI array for integer GCD computation\/},
Technical Report TR-CS-82-11, Department of Computer Science, Australian National University,
Canberra, December 1982.
{[}Appeared in {\em Proc.\ ARITH-7}, IEEE/CS Press, 1985.{]}\\

$[12]$& R.P.~Brent and F.T.~Luk,
Computing the Cholesky factorization using a systolic architecture,
{\em Proceedings of the Sixth Australian Computer Science Conference\/},
Sydney, February 1983, 295--302.\\

$[13]$& R.P.~Brent and F.T.~Luk,
{\em A systolic architecture for the singular value decomposition\/},
Technical Report TR-CS-82-09, Department of Computer Science, Australian National University,
Canberra, August 1982.\\

$[14]$& R.P.~Brent and F.T.~Luk,
{\em A systolic architecture for almost linear-time
solution of the symmetric eigenvalue problem\/},
Technical Report TR-CS-82-10, Department of\linebreak 
Computer Science,
Australian National University, Canberra, August, 1982.\\

$[15]$& R.P.~Brent and F.T.~Luk,
A systolic array for the linear-time solution of Toeplitz systems of equations,
{\em J.~VSLI and Computer Systems\/} 1 (1983), 1--23.\\

$[16]$& R.P.~Brent and F.T.~Luk,
{\em A comparison of several classes of orderings for the Jacobi method\/},
Technical Report, Department of Computer Science, 
Australian National\linebreak 
University, Canberra, to appear.
{[}A revision and extension appeared in
{\em SIAM J.~Sci.~and Stat.~Comput.\/} 6 (1985), 69--84.{]}\\

$[17]$& R.P.~Brent, F.T.~Luk and C.~Van Loan,
{\em Almost linear-time computation of the singular value decomposition using
mesh-connected processors\/},
Report TR-82-528, Department of Computer Science, Cornell University,
Ithaca, November 1982.
{[}A revision appeared in {\em J.~VLSI and Computer Systems}
1, 3 (1983--1985), 242-270.{]}\\

\end{tabular}

\begin{tabular}{rp{147mm}}

$[18]$& B.~Chazelle,
{\em Comptational geometry on a systolic chip\/}, Report CMU-CS-119, Department of
Computer Science, Carnegie-Mellon University, Pittsburgh, April 1982.\\

$[19]$& S.N.~Cole,
Real-time computation by $n$-dimensional iterative arrays of
finite-state\linebreak
machines,
{\em IEEE Trans.~Comp.\/} C-18 (1969), 349--365.\\

$[20]$& G.~Cybenko,
The numerical stability of the Levinson-Durbin algorithm for Toeplitz systems of
equations,
{\em SIAM J.~Sci.~Stat.~Comput.\/} 1 (1980), 303--320.\\

$[21]$& Y.~Dohi, A.~L.~Fisher, H.T.~Kung and L.M.~Monier,
The programmable systolic chip: project overview,
{\em Proceedings of Workshop on Algorithmically Specialized
Computer\linebreak
Organizations\/}, Purdue
University, Indiana, September 1982.\\

$[22]$& P.J.~Eberlein and J.~Boothroyd,
Solution to the eigenproblem by a norm reducing Jacobi type method,
in $[61]$, 327--338.\\

$[23]$& G.E.~Forsythe and P.~Henrici,
The cyclic Jacobi method for computing the principal values of a complex matrix,
{\em Trans.~Amer.~Math.~Soc.\/} 94 (1960), 1--23.\\

$[24]$&M.J.~Foster and H.T.~Kung, The design of special-purpose VLSI chips,
{\em IEEE Computer\/} 13, 1 (January 1980), 26--40.\\

$[25]$& W.M.~Gentleman and H.T.~Kung,
Matrix triangularization by systolic arrays,
{\em Proceedings of SPIE, Vol.~298: Real-Time Signal Processing IV\/},
Society of Photo-Optical\linebreak
Instrumentation Engineers, Bellingham, Washington, 1981.\\

$[26]$& E.R.~Hanson, On cyclic Jacobi methods, 
{\em J.~SIAM\/} 11 (1963), 448--459.\\

$[27]$& L.S.~Haynes, R.L.~Lau, D.P.~Siewiorek and D.W.~Mizell,
A survey of highly parallel computing,
{\em IEEE Computer\/} 15, 1 (January 1982), 9--24.\\

$[28]$& D.E.~Heller and I.C.F.~Ipsen,
Systolic networks for othogonal equivalence transformations and their applications,
{\em Proceedings of 1982 Conference on Advanced Research in VLSI\/}, MIT, Cambridge,
Massachusetts, 113--122.\\

$[29]$& F.C.~Hennie,
{\em Iterative Arrays of Logical Circuits\/}, MIT Press, 1961.\\

$[30]$& C.G.J.~Jacobi,
\"{U}ber ein leichtes Verfahren
die in der Theorie der Sakularstorungen vorkommenden
Gleichungen numerisch aufzulosen,
{\em J.~Reine Angew.~Math.\/} 30 (1846), 51--95.\\

$[31]$& T.~Kailath, A.~Vieira and M.~Morf,
Inverses of Toeplitz operators, innovations, and orthogonal polynomials,
{\em SIAM Review\,} 20 (1978), 106--119.\\

$[32]$& D.E.~Knuth,
{\em The Art of Computer Programming, Vol.~2: Seminumerical Algorithms\/},
2nd edition, Addison-Wesley, Reading, Massachusetts, 1981.\\

$[33]$& R.H.~Kuhn,
Yield enhancement by fault tolerant systolic arrays,
in $[40]$, 145--152.\\

$[34]$& H.T.~Kung,
The structure of parallel algorithms,
in {\em Advances in Computers\/}, Vol.~19, Academic Press, New York, 1980, 65--112.\\

$[35]$& H.T.~Kung,
Why systolic architectures?,
{\em IEEE Computer\/} 15, 1 (January 1982), 37--46.\\

$[36]$& H.T.~Kung and P.L.~Lehman,
Systolic (VLSI) arrays for relational database operations,
{\em Proceedings of ACM-SIGMOD 1980 International Conference on Management of Data\/},
May 1980, 105--116.\\

$[37]$& H.T.~Kung and C.E.~Leiserson,
Systolic arrays (for VLSI),
in $[47]$, Section~8.3, 271--292.\\

$[38]$& H.T.~Kung and S.W.~Song,
A systolic 2-D convolution chip,
in {\em Multicomputers and Image Processing\/} (K.~Preston and L.~Uhr, editors),
Academic Press, New York, 1982, 373--384.\\

$[39]$& H.T.~Kung, R.F.~Sproull and G.L.~Steele (editors),
{\em VLSI Systems and Computations\/}, Computer Science Press, Maryland, 1981.\\

$[40]$& S.Y.~Kung (editor),
{\em Proceedings of USC Workshop on VLSI and Modern Signal Processing\/},
University of Southern California, Los Angeles, November 1982.\\

$[41]$& S.Y.~Kung,
Impact of VLSI on modern signal processing, in $[40]$, 123--132.\\

$[42]$& S.Y.~Kung and Y.H.~Hu,
Fast and parallel algorithms for solving Toeplitz systems,
{\em Proceedings of International Symposium on Mini and Microcomputers in Control and
Measurement\/},
San Francisco, California, May 1981, 163--168.\\

\end{tabular}

\begin{tabular}{rp{147mm}}
$[43]$& C.E.~Leiserson,
{\em Area-efficient VLSI Computation\/},
Report CMU-CS-82-108, Department of Computer Science, Carnegie-Mellon University,
Pittsburgh, October 1981.
{[}Published by MIT Press, 1983.{]}\\

$[44]$& N.~Levinson,
The Wiener RMS (root-mean-square) error criterion in filter design and prediction,
{\em J.~Math.~Phys.\/} 25 (1947), 261--278.\\

$[45]$& R.F.~Lyon,
A bit-serial VLSI architectural methodology for signal processing,
in {\em VLSI 81\/} (J.P.~Gray, editor), Academic Press, New York, 1981, 131--140.\\

$[46]$& F.J.~MacWilliams and N.J.~Sloane,
{\em The Theory of Error Correcting Codes\/}, North-Holland, Amsterdam, 1977.\\

$[47]$&C.A.~Mead and L.A.~Conway,
{\em Introduction to VLSI Systems\/},
Addison-Wesley, Reading, Massachusetts, 1980.\\

$[48]$& M.~Morf,
Doubling algorithms for Toeplitz and related equations,
{\em Proceedings of IEEE International Conference on Acoustics, Speech and Signal
Processing\/},
IEEE, New York, 1980, 954--959.\\

$[49]$& J.G.~Nash and G.R.~Nudd,
Concurrent VLSI architectures for two-dimensional signal processing systems,
in $[40]$, 166--173.\\

$[50]$& H.~Rutishauser,
The Jacobi method for real symmetric matrices,
in $[61]$, 202--211.\\

$[51]$& A.H.~Sameh,
On Jacobi and Jacobi-like algorithms for a parallel computer,
{\em Math.~Comp.\/} 25 (1971), 579--590.\\

$[52]$& R.~Schreiber,
Systolic arrays for eigenvalue computation,
{\em Proceedings of SPIE Symp.~East 1982, Vol.~341: Real-time Signal Processing V\/},
Society of Photo-Optical Instrumentation Engineers, %
1982.\\

$[53]$& J.~Schur,
\"{U}ber Potenzreihen, die im Innern des Einheitskreises beschr\"{a}nkt sind,
{\em J.~Reine Angew.~Math.\/} 147 (1917), 205--232.\\

$[54]$& H.~Sexton, M.~Shensa and J.~Speiser,
Remarks on a displacement-rank inversion method for Toeplitz systems,
{\em Linear Algebra and its Applications\/} 45 (1982), 127--130.\\

$[55]$& L.~Snyder,
Introduction to the configurable, highly parallel computer,
{\em IEEE Computer\/} 15, 1 (January 1982), 47--56.\\

$[56]$& D.R.~Sweet,
{\em Numerical Methods for Toeplitz Matrices\/},
Ph.D.~thesis, Department of\linebreak 
Computing Science, University of Adelaide, May 1982.\\

$[57]$&J.J.~Symanski,
Progress on a systolic processor implementation,
{\em Proceedings of SPIE Symp.~East 1982, Vol.~341: Real-time Signal Processing V\/},
Society of Photo-Optical Instrumentation Engineers, %
1982.\\

$[58]$& G.~Szeg\"{o},
{\em Orthogonal Polynomials\/},
3rd edition, AMS Colloquium Publication, Vol.~23, American Math.~Society,
Providence, Rhode Island, 1967.\\

$[59]$&J.H.~Wilkinson,
Note on the quadratic convergence of the cyclic Jacobi process,\linebreak
{\em Numer.~Math.\/} 4 (1962), 296--300.\\

$[60]$& J.H.~Wilkinson,
{\em The Algebraic Eigenvalue Problem\/},
Clarendon Press, Oxford, 1965.\\

$[61]$& J.H.~Wilkinson and C.~Reinsch (editors),
{\em Handbook for Automatic Computation, Vol.~2: Linear Algebra\/},
Springer-Verlag, Berlin, 1971.\\

$[62]$& S.~Zohar,
Toeplitz matrix inversion:  The algorithm of W.F.~Trench,
{\em J.~ACM\/} 16 (1969), 592--601.
\end{tabular}

\end{document}